\documentclass[pdflatex,sn-mathphys-num]{sn-jnl}%

\usepackage{graphicx}%
\usepackage{multirow}%
\usepackage{amsmath,amssymb,amsfonts}%
\usepackage{amsthm}%
\usepackage[title]{appendix}%
\usepackage{xcolor}%
\usepackage{textcomp}%
\usepackage{manyfoot}%
\usepackage{booktabs}%
\usepackage{algorithm}%
\usepackage{algorithmicx}%
\usepackage{microtype}%
\usepackage{algpseudocode}%
\usepackage{listings}%
\usepackage{natbib}
\usepackage{cleveref}

\newtheorem{example}{Example}

\usepackage{hyperref}
\usepackage{tikz}
\usepackage{tikz-3dplot}
\usetikzlibrary{shapes.geometric, arrows.meta, positioning, fit, backgrounds, calc, shadows.blur}

\usepackage{bm}%
\usepackage{enumerate}%
\usepackage{float}%

\graphicspath{{plots/}}

\tdplotsetmaincoords{60}{120}

\definecolor{modulecolor}{RGB}{66, 133, 244}    %
\definecolor{datacolor}{RGB}{150, 150, 150}     %
\definecolor{maincolor}{RGB}{234, 67, 53}       %

\definecolor{newtext}{RGB}{0, 138, 62}          %
\definecolor{oldtext}{RGB}{136, 51, 187}        %

\numberwithin{equation}{section}

\begin{document}

\title[IOD of NEOs from VSAs with NNs]{Comparative analysis of Neural Networks approaches for Initial Orbit Determination of Near-Earth Objects}

\author*[1]{\fnm{Francesco} \sur{Geroni}}\email{francesco.geroni@phd.unipi.it}

\author[1]{\fnm{Roberto} \sur{Paoli}}

\author[2]{\fnm{Riccardo} \sur{Massidda}}

\author[1]{\fnm{Giacomo} \sur{Tommei}}

\affil*[1]{\orgdiv{Department of Mathematics}, \orgname{University of Pisa}, \orgaddress{\city{Pisa}, \country{Italy}}}

\affil[2]{\orgdiv{Department of Computer Science}, \orgname{University of Pisa}, \orgaddress{\city{Pisa}, \country{Italy}}}

\abstract{%
Initial Orbit Determination (IOD) from Very Short Arcs (VSAs) remains one of the most challenging open problems in asteroid surveillance and celestial mechanics. Classical methods require angular observations spanning a sufficient fraction of the orbit, and become ill-conditioned or fail outright
when only a single observing night is available, as is the case for most newly discovered Near-Earth Objects (NEOs). We present a comparative analysis of two Neural Network (NN) models that attack the ranging problem directly: both ingest a triplet of time-tagged angular measurements $(t_i,\alpha_i,\delta_i)$, $i=1,2,3$, from a single VSA and predict the
geocentric range and range-rate $(\rho,\dot\rho)$, thus completing the orbital state vector. The first is a Multi-Layer Perceptron trained on a purely data-driven objective; the second augments it with a physics-informed loss built on the Gauss scalar range ODE, together with a distributional term that
counteracts the collapse onto the degenerate near-observer solution. Both are trained, under an object-disjoint partition, on a sample of the $568\,127$ VSAs of $39\,031$ real NEOs available in the NEODyS-2 catalogue. Each model returns an estimate on every arc by construction, so we assess instead whether that estimate is
dynamically admissible: on a held-out test set of $56\,773$ arcs the
physics-informed model places $85.1\%$ of its predictions inside the admissible region, against $76.5\%$ for the data-driven baseline, whereas Gauss's and Laplace's methods return a solution on only $45.0\%$ and $47.3\%$ of the same arcs and collapse onto the degenerate root in about nine of those
cases out of ten. A stratified analysis over proper motion and true range shows that the physics-informed objective is not uniformly superior to the baseline: it trades a longer error tail for admissibility and for a marked advantage on the fast, nearby arcs that are operationally the most relevant. We further report distribution-free prediction intervals calibrated per proper-motion bin, and quantify the degeneracy that a single night cannot remove.}

\keywords{Initial Orbit Determination, Near-Earth Objects, Very Short Arcs, Neural Networks, Physics-Informed Neural Networks, Admissible region, Conformal prediction}

\maketitle

\section{Introduction}%
\label{sec1}

The discovery rate of Near-Earth Objects (NEOs) has increased dramatically due to modern survey programs such as the Catalina Sky Survey and the Vera C. Rubin Observatory, generating thousands of new tracklets per night. A significant fraction of these observations spans only VSAs, typically lasting a few hours within a single observing session, thus providing limited orbital information~\cite{milani2004virtual}.
Classical IOD methods, such as those developed by Gauss and Laplace~\cite{danby1992}, require observations that cover a sufficiently large portion of the orbit. When applied to VSAs, these methods often become ill-conditioned or fail entirely due to the limited curvature information available in the observations~\cite[\S9.1]{milani2010theory}\cite{milani2004orbit}. As a consequence, many newly detected objects are lost before follow-up observations can be scheduled, posing a serious challenge for planetary defense efforts~\cite{milani2004virtual}.
Data-driven Machine Learning (ML) methods, such as deep Neural Networks (NNs), have recently gained traction in asteroid dynamics and astrodynamics more broadly~\cite{carruba2022machine,izzo2022selected}.
Physics-Informed Neural Networks (PINNs)~\cite{raissi2019physics,cuomo2022scientific} are a particularly promising paradigm: they learn from data while simultaneously satisfying governing equations through additional loss terms, regularizing the solution space with physical priors.
NNs have been explored in several astrodynamics contexts~\cite{izzo2024geodesy,schiassi2022physics}, and network-based approaches to orbit determination itself have begun to appear~\cite{scorsoglio2023physic}. Their application to the VSA regime specifically, where the observational geometry is highly degenerate and a single night supplies no curvature to exploit, remains largely unexplored. In this work, we propose a simple Multi-Layer Perceptron (MLP) and a PINN architecture that take as input a triplet of angular observations from a single night and directly predict $(\rho,\dot\rho)$. In particular, the physics-informed component enforces the Gauss scalar range ODE, the projection of the two-body EOM onto the line of sight, as a Raissi-style physics-guided loss~\cite{raissi2019physics}, requiring no ground-truth orbital elements at inference time. The MLP and the PINN performances are then compared with each other and with the ones of classical methods through different metrics. \\
The remainder of this paper is organized as follows.
Section~\ref{sec:IODproblem} reviews the mathematical framework of the IOD problem, including the attributable formalism and the classical methods.
Section~\ref{sec:Dataset} presents the dataset construction and pre-processing. Section~\ref{sec:ModelArchitecture} describes the architecture of the MLP and PINN models and the training strategy, including the design of the loss functions.
In Section~\ref{sec:Comparison_metrics} we explain the different metrics used to compare the performance of the NN models investigated in this work.
Section~\ref{sec:Results} reports training experiments and preliminary results, including comparisons between the two NN models and against classical IOD methods.
Section~\ref{sec:Conclusion_FutureWork} concludes discussing the results with current limitations and future directions.

\section{The Problem of Orbit Determination}%
\label{sec:IODproblem}

\begin{figure}
  \centering
  \resizebox{0.7\linewidth}{!}{%
      \begin{tikzpicture}[scale=1, tdplot_main_coords, line join=round, line cap=round,
                          every node/.append style={font=\tiny,black}]
        \tdplotsetmaincoords{60}{120}
        \def\rSphere{2.5}\def\thetaO{60}\def\phiO{45}\def\rSun{4}\def\thetaSun{270}
        \pgfmathsetmacro{\SunX}{\rSun*cos(\thetaSun)}
        \pgfmathsetmacro{\SunY}{\rSun*sin(\thetaSun)}
        \pgfmathsetmacro{\SunZ}{0}
        \pgfmathsetmacro{\ObjX}{\rSphere*cos(\thetaO)*cos(\phiO)}
        \pgfmathsetmacro{\ObjY}{\rSphere*sin(\thetaO)*cos(\phiO)}
        \pgfmathsetmacro{\ObjZ}{\rSphere*sin(\phiO)}
        \draw[orange,->] (0,0,0) -- (\SunX,\SunY,\SunZ) node[midway, below=-2pt] {$r_{\oplus}$};
        \draw[violet,->] (\SunX,\SunY,\SunZ) -- (\ObjX,\ObjY,\ObjZ) node[midway, above=-2pt] {$r$};
        \tdplottransformmainscreen{0}{0}{0}
        \begin{scope}[tdplot_screen_coords, shift={(\tdplotresx,\tdplotresy)}]
          \draw[gray!40!white,opacity=0.6] (0,0) circle (\rSphere);
          \draw[blue!40!white,opacity=0.7,thick] (0,0) ellipse ({\rSphere} and {\rSphere*0.4});
          \draw[gray!50!white,opacity=0.4] (0,0) ellipse ({\rSphere*cos(30)} and {\rSphere*cos(30)*0.3});
          \draw[gray!50!white,opacity=0.4] (0,0) ellipse ({\rSphere*cos(60)} and {\rSphere*cos(60)*0.15});
          \draw[gray!50!white,opacity=0.4] (0,0) ellipse ({\rSphere*0.8} and \rSphere);
          \draw[gray!50!white,opacity=0.4] (0,0) ellipse ({\rSphere*0.5} and \rSphere);
          \draw[gray!50!white,opacity=0.4] (0,0) ellipse ({\rSphere*0.2} and \rSphere);
        \end{scope}
        \draw[thick,->] (0,0,0) -- (\ObjX,\ObjY,\ObjZ) node[midway, above=-1pt] {$\rho$};
        \pgfmathsetmacro{\ProjX}{\rSphere*cos(\thetaO)*cos(\phiO)}
        \pgfmathsetmacro{\ProjY}{\rSphere*sin(\thetaO)*cos(\phiO)}
        \draw[dotted,black,thick] (\ObjX,\ObjY,\ObjZ) -- (\ProjX,\ProjY,0);
        \tdplottransformmainscreen{\ProjX}{\ProjY}{0}
        \begin{scope}[tdplot_screen_coords]
          \fill[black,opacity=0.8] (\tdplotresx,\tdplotresy) circle (0.05);
        \end{scope}
        \draw[dotted,black,thick,->] (0,0,0) -- (\ProjX,\ProjY,0);
        \begin{scope}[canvas is xy plane at z=0]
          \draw[red,thick,->] (0,0) ++(0:1.3) arc[start angle=0, end angle=\thetaO, radius=1.3];
          \node[red, anchor=south] at ($(0,0)+(30:2.0)$) {$\alpha$};
        \end{scope}
        \tdplotsetthetaplanecoords{\thetaO}
        \begin{scope}[tdplot_rotated_coords]
          \draw[green!70!black,thick,->] (0,0,0) ++(90:1.3) arc[start angle=90, end angle=90-\phiO, radius=1.3];
          \node[green!70!black, anchor=south west] at (90-\phiO/2.1:1) {$\delta$};
        \end{scope}
        \draw[->] (0,0,0) -- (2.7,0,0) node[below] {$x$};
        \draw[->] (0,0,0) -- (0,2.7,0) node[below] {$y$};
        \draw[->] (0,0,0) -- (0,0,3.1) node[right] {$z$};
        \tdplottransformmainscreen{0}{0}{0}
        \begin{scope}[tdplot_screen_coords]
          \fill[ball color=blue!60!white,opacity=0.95] (\tdplotresx,\tdplotresy) circle (0.25);
          \node[left=5pt] at (\tdplotresx,\tdplotresy) {Earth};
        \end{scope}
        \tdplottransformmainscreen{\SunX}{\SunY}{\SunZ}
        \begin{scope}[tdplot_screen_coords]
          \fill[ball color=orange!80!yellow,opacity=0.95] (\tdplotresx,\tdplotresy) circle (0.4);
          \node[above=10pt] at (\tdplotresx,\tdplotresy) {Sun};
        \end{scope}
        \tdplottransformmainscreen{\ObjX}{\ObjY}{\ObjZ}
        \begin{scope}[tdplot_screen_coords]
          \fill[ball color=red!70!black,opacity=0.95] (\tdplotresx,\tdplotresy) circle (0.1);
          \node[right=3pt] at (\tdplotresx,\tdplotresy) {NEO};
        \end{scope}
      \end{tikzpicture}}
    \caption{Reference frame in consideration: Earth is positioned at the origin of the celestial sphere, while the Sun is placed on its orbital plane and the body is placed on the sphere surface. Here $\alpha$ represents the \textit{Right Ascension} (RA), $\delta$ is the \textit{Declination} (Dec), $\rho$ is the NEO distance vector (3D), and $r_{\oplus}$ is the Sun distance vector (2D).}
    \label{fig:reference_frame}
\end{figure}
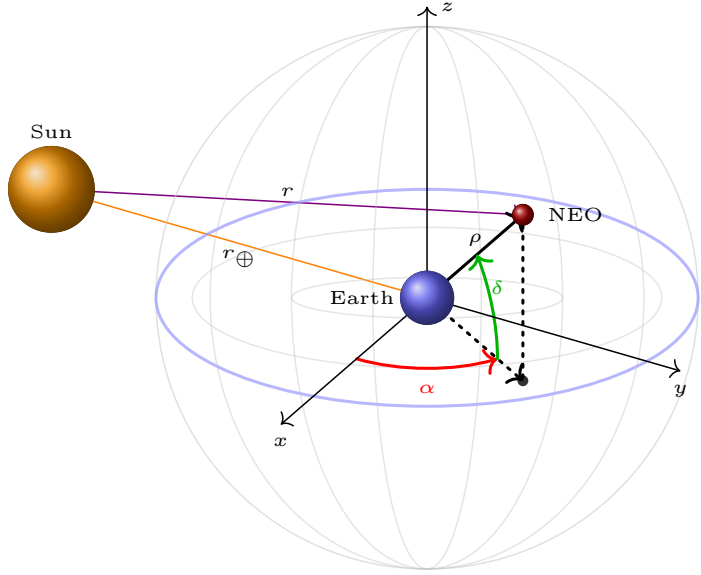

The core mathematical difficulty of IOD lies in the \emph{ranging problem}: from angular observations alone (RA~$\alpha$, Dec~$\delta$) one must recover the geocentric range~$\rho$ and its time derivative~$\dot\rho$ which, together with the observer's known state, complete the geocentric position and velocity of the object (\Cref{fig:reference_frame}). For VSAs this problem is severely ill-conditioned.

\subsection{Problem setting}
Consider a ground-based observer recording $\alpha$ and $\delta$ of a celestial object at times $t_1 < t_2 < t_3$ within a single observing night.
Throughout this work the observer is approximated at the geocentre: the topocentric parallax due to the observatory position on the Earth's surface ($\lesssim 4.3\times10^{-5}$~AU) is neglected~\cite[\S8.3]{milani2010theory}, an approximation adequate everywhere except for extremely close approaches.
The \emph{attributable}~\cite[\S7.6]{milani2010theory} at the central epoch~$t_2$ is the four-dimensional vector
\begin{equation}
    \mathcal{A} = (\alpha,\, \delta,\, \dot\alpha,\, \dot\delta),
\end{equation}
where the angular rates are reconstructed at the middle epoch with a Lagrange 3-point derivative:
\begin{equation}
    \dot\alpha(t_2) \approx
    \frac{(\alpha_2-\alpha_1)}{h_1}\frac{h_2}{h}
    +
    \frac{(\alpha_3-\alpha_2)}{h_2}\frac{h_1}{h}\,, \qquad
    \dot\delta(t_2) \approx
    \frac{\delta_2-\delta_1}{h_1}\frac{h_2}{h}
    +
    \frac{\delta_3-\delta_2}{h_2}\frac{h_1}{h}\,,
    \label{eq:attr_rates}
\end{equation}
with $h_1=t_2-t_1$, $h_2=t_3-t_2$, $h=t_3-t_1$ and
\((\Delta\alpha)=\operatorname{atan2}(\sin\Delta\alpha,\cos\Delta\alpha)\).
The attributable $\mathcal{A}$ is immediately computable from the observations and describes four of the six independent components of the geocentric state. The missing information is the pair $(\rho,\,\dot\rho)$, which completes the six-dimensional state of \emph{attributable orbital elements}~\cite[\S8.3]{milani2010theory}
\begin{equation}
    \mathcal{K} = (\alpha,\, \delta,\, \dot\alpha,\, \dot\delta,\, \rho,\, \dot\rho).
\end{equation}

\subsection{Classical IOD methods}%
\label{subsec:classical}

The analytical methods of Gauss and Laplace are still the most widely used to obtain an initial solution for the ranging problem, completing the orbital state $\mathcal{K}$. Both methods assume two-body Keplerian motion and are limited by the quality of available data and the portion of orbit that they cover, meaning that the resulting estimate may be far from the true solution or that they may not converge at all. The problem is moreover not guaranteed to have a unique answer: the same angular data can admit several distinct orbital solutions, whose computation and occurrence have been studied in detail by Milani et al.~\cite{milani2004asteroid}.
This is because in the VSA regime the curvature information is weak: $\dot\alpha,\, \dot\delta$ become nearly constant, higher-order curvature terms vanish, and the matrix conditioning of the resulting system deteriorates rapidly~\cite[\S6.3, \S9.1]{milani2010theory}\cite{milani2004asteroid}.
The degeneracy of the solutions is also well illustrated by the concept of \emph{admissible region}: the family of possible pairs of $(\rho, \dot\rho)$ that are compatible with the angular observations lie inside a region, defined by suitable admissibility conditions, that cannot be reduced further until more information is collected~\cite{milani2004orbit,milani2004asteroid}.

\section{Dataset \& Preprocessing}%
\label{sec:Dataset}

Having a high-quality dataset is a fundamental prerequisite for the success of any ML model.
We developed a comprehensive acquisition and preprocessing pipeline specifically designed for training NNs on real NEO observations, arranged into VSAs.

\subsection{Dataset Creation}

\begin{figure}
    \centering
    \includegraphics[width=1\textwidth]{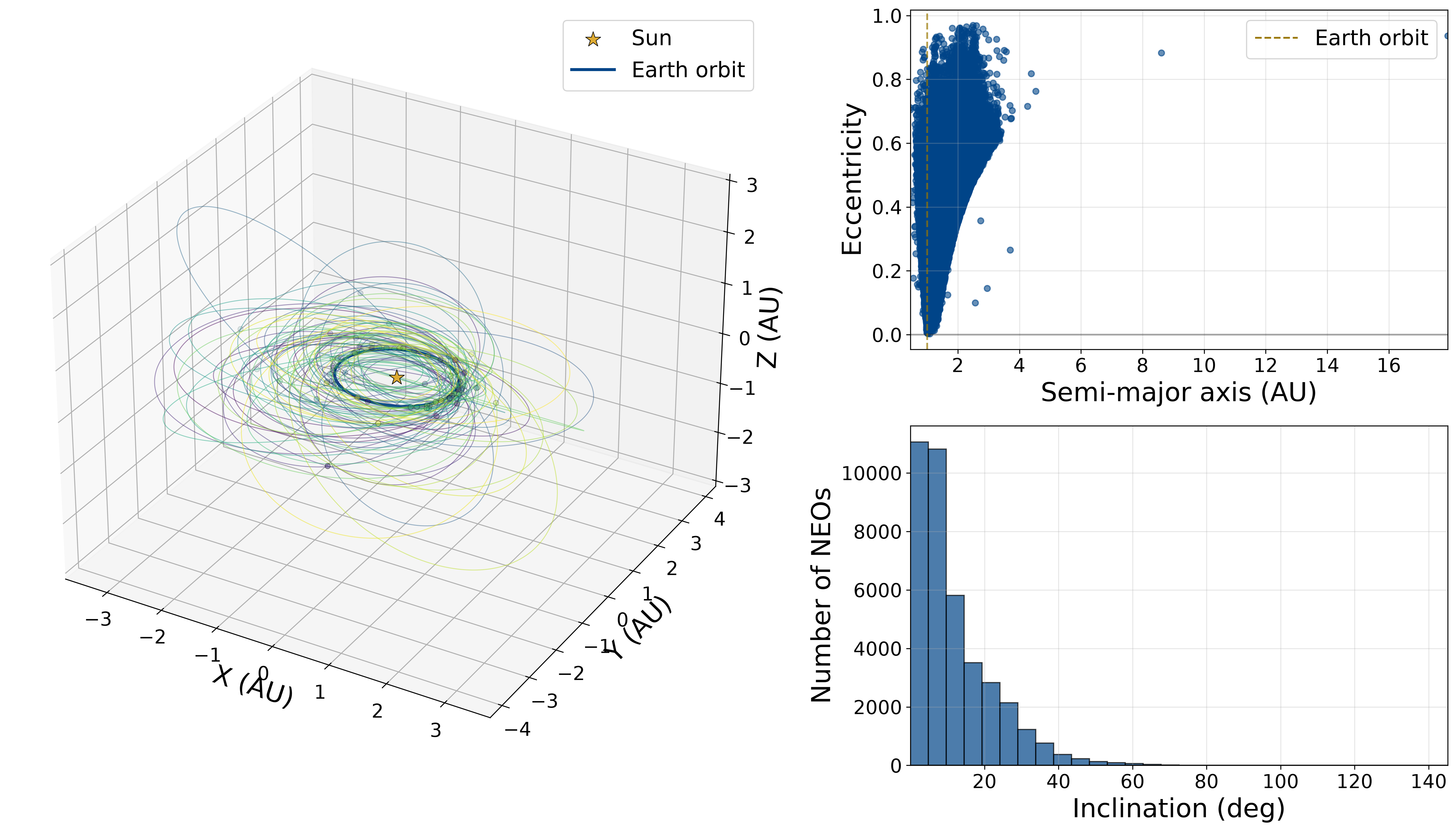}
    \caption{Visualization of the NEO dataset constructed from the NEODyS-2 catalogue. Left: 3D orbital trajectories of a subset of 80 NEOs relative to Earth's orbit and the Sun. Top-right: semi-major axis vs.\ eccentricity, with Earth's orbit marked at 1~AU. Mid-right: inclination distribution histogram of the same samples.}\label{fig:NEO_Dataset_Visualization}
\end{figure}

The pipeline interfaces directly with the NEODyS-2\footnote{\url{https://newton.spacedys.com/neodys/}} portal using a custom-made observational query, implementing a web-scraping and validation system that accesses two distinct data sources for each object:
\begin{enumerate}[1.]
    \item \textbf{Observational files} (\texttt{.rwo}): Time-stamped angular measurements ($\alpha$, $\delta$) from ground-based telescopes, with precise timestamps in Modified Julian Date (MJD). Everything derived from the observations is in the equatorial frame (ICRS). Each record also carries the a-priori uncertainty of the measurement, the Minor Planet Center observatory code and the flag with which NEODyS marks the observation as used or rejected in its own fit;
    \item \textbf{Keplerian elements}: Best-fit classical orbital parameters ($a$, $e$, $i$, $\Omega$, $\omega$, $M$) extracted from the NEODyS-2 HTML orbit tables. These are expressed in the ecliptic frame, and the perifocal rotation $R(\Omega, i, \omega)$ returns vectors in the frame of the elements it is given.
\end{enumerate}
The necessary information is extracted as columns (one entry for each object's VSA) and stored in two different \verb|.csv| files: one for raw observations and one for the \emph{ground truth} (these are the "true" orbital data against which the NN model predictions are compared during the training phase) extracted from the Keplerian elements table. An overall example of the dataset is illustrated in Fig.~\ref{fig:NEO_Dataset_Visualization}. Two selections are applied at ingestion, before any arc is formed: 1) only optical records are kept, consistently with the geocentric approximation, while space-based and radar measurements, which are of a different observable type, are discarded in this work; 2) observations that NEODyS itself flags as rejected outliers in its own orbital fit are dropped.
\\
For each object and observing night with at least three observations, a triplet $(t_1, \alpha_1, \delta_1)$, $(t_2, \alpha_2, \delta_2)$, $(t_3, \alpha_3, \delta_3)$ is selected.
All RA differences use the $\mathrm{atan2}(\sin\Delta\alpha, \cos\Delta\alpha)$ wrapping to handle the $\alpha = 0/2\pi$ discontinuity correctly.
The ground-truth $(\rho, \dot\rho)$ is computed by propagating the Keplerian elements from their reference epoch to the central observation time~$t_2$. We do this by considering the mean anomaly propagation as $M(t_2) = M(\text{epoch}) + n(t_2 - \text{epoch})$, where $n = \sqrt{\mu_\odot/a^3}$ and then solving the geometric inversion to the geocentric reference frame.
Moreover, the targets follow the emission-epoch convention of Milani \& Gronchi~\cite[\S8.3]{milani2010theory}: $\rho \equiv c\,\tau$ is the actual light path, so that $\rho\,\hat{\boldsymbol{\rho}} = \mathbf{r}(t_0) - \mathbf{r_{\oplus}}(\bar t)$ holds exactly, with $t_0 = \bar t - \rho/c$ the epoch of the state and $\bar t$ the epoch of reception. Positions therefore need no correction, whereas the velocity reconstructed from an attributable is retarded by a factor $(1 - \dot\rho/c)$ and is divided by it wherever a heliocentric state is formed. Since $t_0$ depends on $\rho$, the mean anomaly of both prediction and truth is referred to the common reception epoch~$t_2$.
We also applied a minimum quality filter on the considered VSA for numerical stability reasons: triplets with proper motion $\eta \le 10^{-4}$~rad/day ($\approx 0.0057$~deg/day) or total duration $t_3 - t_1 \le 10^{-3}$~day ($\approx 1.44$~min) are discarded, together with those in which two of the three epochs coincide, for which the three-point derivative of Eq.~\eqref{eq:attr_rates} is undefined. Grouping the retained observations by object and night yields $570{,}476$ candidate arcs; the quality filter above discards a further $2{,}349$ of them, so the usable catalogue contains $568{,}127$ VSA arcs drawn from $39{,}031$ NEO objects.

\subsection{Data Preprocessing}\label{subsec:NormScaling}

The network inputs are normalized to a $[-1, 1]$ range feature-by-feature. In particular, the two outer times enter as arc-relative offsets $\Delta t_1 = t_1 - t_2 \le 0$ and $\Delta t_3 = t_3 - t_2 \ge 0$ (in days), normalized symmetrically about zero by the largest half-arc in the dataset, so the $\Delta t_1$ slot lands in $[-1, 0)$ and the $\Delta t_3$ slot in $(0, 1]$.
The central time $t_2$ is retained as the (normalized) absolute MJD, because the physics loss needs it to query the Earth/Sun ephemeris and the evaluation reports against it; it is however not presented to the network, as discussed in \Cref{subsec:data_representation}. The epoch is instead carried by the observer's heliocentric position $\mathbf{r}_\oplus(t_2)$, which is appended to the feature vector and normalized like the other inputs.\\
All observations of a given NEO are assigned to the same train/validation/test partition, preventing \emph{data leakage}~\cite{goodfellow2016deep}.
Whole objects are allocated greedily so that the realised sample counts track the target ratios 70\%/20\%/10\%; the partition is therefore object-disjoint.
Earth and Sun barycentric ephemerides are pre-computed and stored as PyTorch tensor caches (\texttt{.pth} files). This allows an efficient reconstruction of the heliocentric state from the predicted $(\rho^{*}, \dot\rho^{*})$ during training and evaluation, without incurring the overhead of repeated ephemeris queries or numerical integrations.

\section{Neural Network for Orbit Determination}%
\label{sec:ModelArchitecture}

We apply two different Neural Network models in order to test and compare their performance in solving the IOD problem from VSA. Since both of them share the same architecture, we treat the PINN as an extension of the MLP. In the following, we are going to explain the theoretical foundations of these models and how we adapted them to deal with this physical application.

\subsection{Multi-Layer Perceptron}
Neural Networks are universal function approximators~\cite{hornik1989multilayer,cybenko1989approximation} that can be used to approximate an unknown differentiable target function $f:\mathcal{X}\to\mathcal{Y}$, where we denote as $\mathcal{X}$ the input space and as $\mathcal{Y}$ the output space. In detail, a feedforward NN with $L$ hidden layers implements a composition of affine transformations and nonlinear activations. Networks with a large number of layers~$L$ are known as \emph{deep} NNs~\cite{goodfellow2016deep}, the simplest of which is the MLP. Formally, each layer $l\in\{1, \ldots, L\}$ of the network computes
\begin{align}
    \mathbf{h}^{(0)} &= \mathbf{x}\\
    \mathbf{h}^{(l)} &= \sigma(\mathbf{W}^{(l)}\mathbf{h}^{(l-1)} + \mathbf{b}^{(l)}),
\end{align}
where $\mathbf{W}^{(l)}$ is the weight matrix, $\mathbf{b}^{(l)}$ is the bias vector, and $\sigma(\cdot)$ is a nonlinear activation function applied element-wise. Common choices for $\sigma$ include the hyperbolic tangent ($\tanh$) and the rectified linear unit (ReLU)~\cite{goodfellow2016deep}. We denote the set of all parameters as $\theta = \{(\mathbf{W}^{(l)}, \mathbf{b}^{(l)})\}_{l=1}^{L}$ and the overall function implemented by the neural network as $f_\theta:\mathcal{X}\to\mathcal{Y}$. The objective of training a neural network is to find the optimal parameters $\theta^\star$ that approximate the target function $f$ as closely as possible, such that
\begin{equation}
    f_{\theta^\star}(\mathbf{x}) \approx f(\mathbf{x}), \quad \forall \mathbf{x} \in \mathcal{X}.
\end{equation}

Given a dataset of $N$ observations $\mathcal{D}=(\mathbf{x}_i, y_i)_{i=1}^N$, supervised training of a neural network requires minimizing a loss function $\mathcal{L}_{\mathcal{D}}(\theta)$ that quantifies the distance between the model predictions~$f_\theta(x)$ and the ground-truth outputs~$y$ in the dataset.
Formally,
\begin{equation}
    \theta^\star = \arg\min_\theta \mathcal{L}_{\mathcal{D}}(\theta)
    = \arg\min_\theta \frac{1}{N} \sum_{i=1}^{N} \ell(f_\theta(\mathbf{x}_i), y_i),
\end{equation}
where $\ell$ is a suitable loss function, such as mean squared error (MSE) for regression tasks or cross-entropy for classification tasks. In practice, we solve the optimization problem via stochastic gradient descent by adjusting the parameters over multiple \emph{epochs} (or \emph{generations}), as standard in the machine learning literature~\cite{goodfellow2016deep,rumelhart1986learning}.

\subsection{Physics-Informed Neural Networks}%
\label{subsec:pinn}

While deep NNs have shown remarkable success in approximating complex functions from data~\cite{goodfellow2016deep}, purely data-driven models can struggle to extrapolate beyond the training distribution and may produce physically inconsistent predictions~\cite{karniadakis2021physics}. A PINN, introduced by \citet{raissi2019physics}, addresses this limitation by augmenting the standard data-driven loss with physics-based penalties.
In particular, we consider systems described by a known PDE of the form
\begin{equation}%
\label{eq:pde_general}
    u_t(t, \mathbf{x}) + \mathcal{N}[u](t,\mathbf{x}) = 0,
\end{equation}
where $u_t(t,\mathbf{x})$ is the time derivative of the unknown solution function $u(t,\mathbf{x})$ and $\mathcal{N}[u](t,\mathbf{x})$ is a possibly nonlinear known differential operator, also expressed as a function of one or more partial derivatives of the solution $u_{\theta}$ of the system.\\
Given a dataset of observations $\mathcal{D}$, the goal of a PINN is to learn a function $u_\theta(t,\mathbf{x})$ that approximates the unknown solution $u(t,\mathbf{x})$, while also satisfying the governing PDE. Therefore, the training process involves minimizing a composite loss function that combines data-fitting and physics-informed terms. Formally,
\begin{equation}
\label{eq:pinn_general_loss}
    \mathcal{L}(\theta) = \mathcal{L}_{\mathcal{D}}(\theta) + \mathcal{L}_{\mathrm{PDE}}(\theta)
\end{equation}
where $\mathcal{L}_{\mathcal{D}}(\theta)$ measures the discrepancy between the network's predictions and the observed data, as in a standard deep neural network, and $\mathcal{L}_{\mathrm{PDE}}(\theta)$ quantifies the violation of the PDE constraints (in the following we are going to use the symbol $\partial_k$ instead of $\frac{\partial}{\partial k}$).
In practice, given a set of $M$ collocation points $(t_j, \mathbf{x}_j)_{j=1}^M$ in the domain of interest, the physics-informed loss is computed as
\begin{equation}
    \mathcal{L}_{\mathrm{PDE}}(\theta) = \frac{1}{M} \sum_{j=1}^{M} \ell( \partial_t u_\theta (t_j, \mathbf{x}_j) + \mathcal{N}[u_\theta](t_j, \mathbf{x}_j), 0),
\end{equation}
where $\ell$ is again a suitable loss function such as the mean squared error. The derivative terms of the neural network $u_\theta$ contained in $\mathcal{L}_{\mathrm{PDE}}$ are computed using automatic differentiation through the network's computational graph in most modern deep learning frameworks~\cite{paszke2019pytorch,abadi2016tensorflow}. Notably, automatic differentiation enables exact gradient calculations without the need for finite-difference approximations~\cite{raissi2019physics}.

\begin{example}
For clarity, before introducing our model for IOD, we report the example used by \citet{raissi2019physics} to introduce the PINN framework on Burgers' equation:
\begin{equation}
\begin{split}
    \partial_t u + \mathcal{N}[u]&= 0, \\
    \partial_t u + u \partial_x u - \nu \, \partial_{xx} u &= 0;
\end{split}
\end{equation}
where $u(t,x)$ is the unknown solution and $\nu$ is a constant. Therefore, given a NN $u_\theta(t,x)$ that approximates $u(t,x)$, the physics-informed loss term is
\begin{equation}
    \mathcal{L}_{\mathrm{PDE}}(\theta) = \frac
{1}{M} \sum_{j=1}^{M} \ell(\partial_t u_\theta(t_j, x_j) + u_\theta(t_j, x_j) \partial_x u_\theta(t_j, x_j) - \nu \partial_{xx} u_\theta (t_j, x_j), 0).
\end{equation}
\end{example}

\subsubsection{PINN Loss Function \texorpdfstring{$\mathcal{L}(\theta)$}{L(theta)} for IOD}%
\label{subsec:loss}

In our model the total loss is a weighted combination of different components, each capturing a specific aspect of the problem:
\begin{equation}
    \mathcal{L}(\theta) = \underbrace{\lambda_{\mathrm{res}}\,\mathcal{L}_{\mathrm{res}}(\theta) + \lambda_{\mathcal{W}}\,\mathcal{L}_{\mathcal{W}}(\theta)}_{\text{data}} + \underbrace{\lambda_{\mathrm{eom}}\,\mathcal{L}_{\mathrm{eom}}(\theta) + \lambda_{\mathrm{kin}}\,\mathcal{L}_{\mathrm{kin}}(\theta)}_{\text{physics}}\,,
    \label{eq:total_loss}
\end{equation}
each term being discussed in the following subsections. The first two terms are data-driven losses, while the last two are physics-informed losses. The data-driven losses measure how well the model fits the training data, while the physics-informed losses enforce the physical constraints of the problem.
The \emph{global weights} $\bm\lambda = (\lambda_{\mathrm{res}},\,\lambda_{\mathrm{eom}}, \,\lambda_{\mathrm{kin}}, \,\lambda_{\mathcal{W}})$ balance the contributions of the four loss functions, as is standard in the PINN literature~\cite{wang2023expert}. In the experiments reported in this work they are kept fixed at $\bm\lambda = (1.0,\,0.30,\,0.21,\,0.05)$. It should be stressed that these values are scale factors and not a measure of the influence of each term, since the four components live on very different intrinsic scales.\\
The two physics weights are not equal because their residuals are not on the same scale: with the per-arc normalisation of Eq.~\eqref{eq:f_eom} they enter the objective with comparable magnitude, so $\lambda_{\mathrm{eom}}$ and $\lambda_{\mathrm{kin}}$ are set to equalise their contribution rather than their nominal coefficient, which is the compensation of gradient-scale imbalance recommended in~\cite{wang2021understanding,wang2023expert}. Conversely $\lambda_{\mathcal{W}}$ is small because $\mathcal{L}_{\mathcal{W}}$ is a $\mathcal{W}_1$ distance, whose gradient is scale-free and therefore does not decay as the fit tightens; a larger value would saturate the global gradient clip and drag the other three terms down with it.

\paragraph{Residual loss \texorpdfstring{$\mathcal{L}_{\mathrm{res}}$}{Lres}}
First, we compute the residual loss that evaluates the predictive capabilities of the model. Given a value $d$, we employ the $\operatorname{logcosh}$ loss function, as in
\begin{equation}
   \ell(d)
   =
   \log[\cosh(d)].
\end{equation}
From now on, $\ell$ represents a $\mathrm{logcosh}$ form that satisfies $\ell \approx d^2/2$ for $|d| \ll 1$ and $\ell \approx |d|$ for $|d| \gg 1$, providing quadratic sensitivity near zero and linear robustness to outliers.\\
The first term of Eq.~\eqref{eq:total_loss} is a simple residual function that measures the size of the model's estimation errors and corresponds to one of the data-driven components of the total loss function $\mathcal{L}(\theta)$. Given the network outputs $(\rho^{*}, \dot\rho^{*})$ and the ground truth $(\rho, \dot\rho)$, we define

\begin{equation}
    \mathcal{L}_{\mathrm{res}} =
    \frac{1}{2}\Biggl[
    \frac{1}{N}
    \sum_{i=1}^{N}
    \ell(\rho_i - \rho^{*}_i)
    +
    \frac{1}{N}
    \sum_{i=1}^{N}
    \ell(\dot\rho_i - \dot\rho^{*}_i)
    \Biggr]
    \label{eq:loss_res}
\end{equation}
where $N$ is the total number of evaluated samples (in the batch). Both residuals are evaluated in the normalized output space (Section~\ref{subsec:NormScaling}).

\paragraph{Physics contributions \texorpdfstring{$\mathcal{L}_{\mathrm{eom}}$ \& $\mathcal{L}_{\mathrm{kin}}$}{Leom\&Lkin}}
Our model implements the PINN paradigm by enforcing the \emph{Gauss scalar range ODE} as the physics-informed constraint. We assume that the Earth moves on a Keplerian two-body orbit around the Sun (\Cref{fig:reference_frame}), i.e.,
\begin{equation}
    \ddot{\mathbf{r}}_{\oplus}=-\dfrac{\mu_{\odot}}{r_{\oplus}^3}\mathbf{r}_{\oplus}.
\end{equation}
The NEO is therefore under the gravitational attraction of the Sun and of the direct pull of the Earth, that is
\begin{equation}\label{ddotr_dyn_e}
        \ddot{\mathbf{r}}=-\dfrac{\mu_{\odot}}{r^3}\mathbf{r}-\dfrac{\mu_\oplus}{\rho^2}\hat{\boldsymbol{\rho}},
\end{equation}
where $\mu_{\odot}=\mathcal{G}m_\odot$ and $\mu_\oplus=\mathcal{G}m_\oplus$, while $\mathcal{G}$ is the gravitational constant and $m_\odot$ and $m_\oplus$ are the masses of the Sun and the Earth, respectively.
Then, we use the fact that $\mathbf{r}=\mathbf{r}_{\oplus}+\rho\hat{\boldsymbol{\rho}}$ to rewrite $\ddot{\mathbf{r}}$.
The unit vector $\hat{\boldsymbol{\rho}}$ is a function of the $\alpha$ and $\delta$ of the observed tracklet. In particular one has that
\begin{equation*}
    \hat{\boldsymbol{\rho}}=\begin{pmatrix}
        \cos \delta  \cos \alpha\\
        \cos \delta  \sin \alpha\\
        \sin \delta
    \end{pmatrix}
\end{equation*}
It is useful to compute the partial derivatives of $\hat{\boldsymbol{\rho}}$ with respect to the angular coordinates, which are
    \[ \partial_\alpha \hat{\boldsymbol{\rho}} = \cos \delta \begin{pmatrix}
        -\sin \alpha\\
        \cos \alpha\\
        0
    \end{pmatrix}, \quad
    \partial_\delta \hat{\boldsymbol{\rho}} = \begin{pmatrix}
        -\sin \delta \cos \alpha\\
        -\sin\delta\sin \alpha\\
        \cos \delta
    \end{pmatrix}. \]
We remark that $\|  \partial_\alpha \hat{\boldsymbol{\rho}}\|=\cos \delta$ and $\| \partial_\delta \hat{\boldsymbol{\rho}}\|=1$. Moreover the three vectors $\hat{\boldsymbol{\rho}},\,\partial_\alpha \hat{\boldsymbol{\rho}}/\cos\delta$ and $\partial_\delta \hat{\boldsymbol{\rho}}$ provide a local orthonormal frame on the unit sphere. Therefore, the derivative of $\hat{\boldsymbol{\rho}}$ with respect to time is given by
\begin{equation}
    \dfrac{d \hat{\boldsymbol{\rho}}}{dt}=\partial_{\alpha}\hat{\boldsymbol{\rho}}\,\dot{\alpha}+\partial_{\delta}\hat{\boldsymbol{\rho}}\,\dot{\delta}.
\end{equation}
As a consequence, one has that
\begin{equation}
\dot{\mathbf{r}}=\dot{\mathbf{r}}
_{\oplus}+\dot{\rho}\hat{\boldsymbol{\rho}}+\rho\left(\dot{\alpha}\,\partial_{\alpha}\hat{\boldsymbol{\rho}}+\dot{\delta}\,\partial_{\delta}\hat{\boldsymbol{\rho}}\right).
\end{equation}
To compute higher-order derivatives it is necessary to compute the second derivatives of $\hat{\boldsymbol{\rho}}$ with respect to the coordinates, i.e.:

\[  \partial_{\alpha\alpha}^2\hat{\boldsymbol{\rho}}=\cos \delta\,\begin{pmatrix}
        - \cos \alpha\\
        - \sin \alpha\\
        0
    \end{pmatrix}; \quad
    \partial_{\alpha\delta}^2\hat{\boldsymbol{\rho}}=\sin \delta \,\begin{pmatrix}
        \sin \alpha\\
        -\cos \alpha\\
        0
    \end{pmatrix}; \quad
        \partial_{\delta\delta}^2\hat{\boldsymbol{\rho}}=\begin{pmatrix}
        -\cos \delta \cos \alpha\\
        -\cos \delta \sin \alpha\\
        -\sin \delta
    \end{pmatrix}=-\hat{\boldsymbol{\rho}}.\]

We note that $\|\partial_{\alpha\alpha}^2\hat{\boldsymbol{\rho}}\|=\cos\delta$, $\|\partial_{\alpha\delta}^2\hat{\boldsymbol{\rho}}\|=|\sin\delta|$ and $\|\partial_{\delta\delta}^2\hat{\boldsymbol{\rho}}\|=1$. The scalar products of the line-of-sight direction $\hat{\boldsymbol{\rho}}$ with the second-derivative vectors are

\begin{equation*}
    \hat{\boldsymbol{\rho}}\cdot \partial_{\alpha\alpha}^2 \hat{\boldsymbol{\rho}}=-\cos^2\delta, \quad \hat{\boldsymbol{\rho}}\cdot \partial_{\alpha\delta}^2 \hat{\boldsymbol{\rho}}=0,\quad \hat{\boldsymbol{\rho}}\cdot \partial_{\delta\delta}^2 \hat{\boldsymbol{\rho}}=-1.
\end{equation*}
Therefore we obtain
\begin{align}\label{ddotr_par}
\ddot{\mathbf{r}}=\ddot{\mathbf{r}}_{\oplus}+&\ddot{\rho}\,\hat{\boldsymbol{\rho}}\,+2\dot{\rho}\left(\dot{\alpha}\,\partial_{\alpha}\hat{\boldsymbol{\rho}}+\dot{\delta}\,\partial_{\delta}\hat{\boldsymbol{\rho}}\right)\nonumber\\
+&\rho\left(\dot{\alpha}^2\,\partial^2_{\alpha\alpha}\hat{\boldsymbol{\rho}}+2\,\dot{\alpha}\dot{\delta}\,\partial^2_{\alpha\delta}\hat{\boldsymbol{\rho}}- \dot{\delta}^2\,\hat{\boldsymbol{\rho}}+\ddot{\alpha}\,\partial_{\alpha}\hat{\boldsymbol{\rho}}+\ddot{\delta}\,\partial_{\delta}\hat{\boldsymbol{\rho}}\right).
\end{align}
By equating Eq.~\eqref{ddotr_par} to Eq.~\eqref{ddotr_dyn_e}, rearranging the terms and projecting along the line of sight $\hat{\boldsymbol{\rho}}$ one obtains the following scalar equation:
\begin{equation}
    \ddot{\rho} = \rho\left(\dot{\alpha}^2\,\cos^2\delta + \dot{\delta}^2\right) + \mu_{\odot}\,\mathbf{r}_{\oplus}\cdot\hat{\boldsymbol{\rho}}\left(\dfrac{1}{r_{\oplus}^3}-\dfrac{1}{r^3}\right) -\dfrac{\mu_{\odot}}{r^3}\rho - \dfrac{\mu_\oplus}{\rho^2}.
    \label{PINN_DDOTRHO}
\end{equation}
Two further scalar equations follow from projecting onto $\partial_\alpha\hat{\boldsymbol{\rho}}$ and $\partial_\delta\hat{\boldsymbol{\rho}}$, but they are of a different nature and are not used here since both projections would cancel the $\ddot\rho$ contribution and compare $\ddot\alpha$ and $\ddot\delta$, which are observed quantities, against $\rho$ and $\dot\rho$ algebraically. They are therefore not differential constraints on the network output at all.
Following the general formulation of Eq.~\eqref{eq:pde_general}, since the network output is $u_\theta(\mathbf{x}) = (\rho^*,\,\dot\rho^*)$, the differential condition decomposes into two equations, both scaled to have dimensionless contributions:
\begin{enumerate}
    \item  The \emph{kinematics} enforces consistency between the predicted $\dot\rho^{*}$ and the time derivative of $\rho^{*}$
            \begin{equation}
                \mathcal{R}_{\mathrm{kin}}  = \left( \frac{\mathrm{d}\rho^*}{\mathrm{d}t} - \dot\rho^* \right) \frac{1}{V_c} = 0 \,,
                \label{eq:f_kin}
            \end{equation}
            where $V_c = \sqrt{\mu_\odot/L_c} \approx 1.72\times10^{-2}$~AU/day is the characteristic velocity and $L_c = 1$~AU is the characteristic length.
    \item  The \emph{equation-of-motion} enforces the Gauss scalar range ODE
            \begin{equation}
                \mathcal{R}_{\mathrm{eom}}  = \left( \frac{\mathrm{d}\dot\rho^*}{\mathrm{d}t} - \ddot\rho \right) a_c^{-1} = 0 \,, \quad a_c = \dfrac{\mu_\odot}{L_c^2} + \eta^2 L_c
                \label{eq:f_eom}
            \end{equation}
            where $\ddot\rho$ is the right-hand side of Eq.~\eqref{PINN_DDOTRHO} evaluated at the predicted state, and $a_c$ is a characteristic gravitational acceleration with a calibration on $\eta$, the proper motion of the arc. We chose this normalization logic so that the residuals, whose typical magnitude varies strongly across the population, are neither dominated by the fast tail nor negligible on the rest of it.
\end{enumerate}
The two contributions of Eq.~\eqref{eq:total_loss} are then defined as the mean of the $\mathrm{logcosh}$ of the residuals over the batch:
\begin{equation}
    \mathcal{L}_{\mathrm{kin}} = \frac{1}{N}\sum_{i=1}^{N}
    \ell\!\left(\mathcal{R}_{\mathrm{kin},i}\right)\,,\quad
    \mathcal{L}_{\mathrm{eom}} = \frac{1}{N}\sum_{i=1}^{N}
    \ell\!\left(\mathcal{R}_{\mathrm{eom},i}\right).
    \label{eq:losses_pde}
\end{equation}
For numerical robustness, the angular rates and accelerations reconstructed via Eq.~\eqref{eq:attr_rates} are clipped to a loose physical envelope ($|\dot\alpha|,\,|\dot\delta| \le 1$~rad/day and $|\ddot\alpha|,\,|\ddot\delta| \le 0.5$~rad/day$^2$) and the dimensionless residuals $\mathcal{R}_{\mathrm{kin}},\,\mathcal{R}_{\mathrm{eom}}$ to $\pm100$ before entering Eq.~\eqref{eq:losses_pde}.

\paragraph{Wasserstein loss \texorpdfstring{$\mathcal{L}_{\mathcal{W}}$}{LW}}
The Wasserstein-1 distance ($\mathcal{W}_1$), also known as the \emph{Earth Mover's Distance}, quantifies the minimum transport cost required to reshape one probability distribution into another~\cite{villani2009optimal,peyre2019computational}. $\mathcal{W}_1$ remains finite and provides informative gradients regardless of support overlap, a property that has been shown to stabilize training of NNs on multimodal problems~\cite{arjovsky2017wasserstein,peyre2019computational}.\\
A fundamental challenge we found applying the PINN in this specific context is the mode collapse behaviour: since the mapping from angular observations to $(\rho^{*},\dot\rho^{*})$ is ill-posed and Eqs.~\eqref{eq:f_kin} and~\eqref{eq:f_eom} admit degenerate solutions, the optimization can converge to a collapsed estimate. A distributional term counteracts this by penalising the discrepancy between the empirical distribution of the predictions and that of the targets. For one-dimensional distributions the $\mathcal{W}_1$ distance has the form
\begin{equation}
    \mathcal{W}_1(P, Q) = \int_0^1 \bigl|F_P^{-1}(q) - F_Q^{-1}(q)\bigr|\,\mathrm{d}q\,,
    \label{eq:wasserstein_def}
\end{equation}
where $F_P^{-1}$ and $F_Q^{-1}$ are the quantile functions of the distributions $P$ and $Q$, respectively. This allows the distance to be computed directly from the quantile functions, with no optimal-transport solver~\cite{peyre2019computational}.\\
Our $\mathcal{L}_{\mathcal{W}}$ is defined as a sum of a global and a conditioned part,

\begin{equation*}
    \mathcal{L}_{\mathcal{W}}(\theta) = \mathcal{L}_{\mathcal{W}}^{\mathrm{global}}(\theta) + \mathcal{L}_{\mathcal{W}}^{\mathrm{cond}}(\theta)\,,
\end{equation*}

\begin{enumerate}[1.]
    \item \textbf{Global component $\mathcal{L}_{\mathcal{W}}^{\mathrm{global}}$:} for each output channel, the $\mathcal{W}_1$ is evaluated on a uniform grid of $51$ quantiles, supplemented by matching terms and by one-sided hinges on the $10$th and $90$th percentiles that penalise the prediction falling inside the target band.
    \item \textbf{Conditioned component $\mathcal{L}_{\mathcal{W}}^{\mathrm{cond}}$:} the same $\mathcal{W}_1$ matching applied to $\rho$ within bins of proper motion $\eta = \sqrt{\dot\alpha^2\cos^2\delta + \dot\delta^2}$ and declination sign, together with an ordering constraint that penalizes $\bar\rho^*_{\mathrm{fast}} > \bar\rho^*_{\mathrm{slow}}$, encoding the physical prior that, generally, fast-moving objects are closer.
\end{enumerate}
The complete mathematical derivation is given in Appendix~\ref{appendix:wasserstein}.

\subsection{Feature Representation}%
\label{subsec:data_representation}

The choice of input representation is crucial for the success of a ML model. In our framework, we compose the input vector $\mathbf{x}$ from the three observations, as described in \Cref{subsec:NormScaling}, together with the heliocentric position $\mathbf{r}_\oplus$ of the observer at the reference epoch $t_2$:
\begin{equation}
    \mathbf{x} = \left[\;\underbrace{\Delta t_1, \alpha_1, \delta_1,\; \alpha_2, \delta_2,\; \Delta t_3, \alpha_3, \delta_3}_{\text{observations (8)}}, \underbrace{\mathbf{r}_\oplus(t_2)}_{\text{observer (3)}}\right] \,.
    \label{eq:input_vector}
\end{equation}
The idea is to maximize the length of the available arc window by selecting $t_1$ and $t_3$ as, respectively, the first and the last observation of the object in the VSA.
Consequently, $t_2$ is the observation closest to the middle of the arc (or equal to it in case of only three observations), which is the reference epoch for the output state of the network. The absolute epoch $t_2$ is deliberately not connected to the trunk, so that
\begin{equation}
    \frac{\partial u_\theta}{\partial t_2} \equiv 0
    \label{eq:no_calendar}
\end{equation}
holds exactly, by construction of the architecture rather than as a property the optimizer might or might not learn. The motivation is that $t_2$ spans nearly four decades of survey history, and a raw calendar axis is a channel along which a network can manufacture a secular trend in $\rho$ that has no dynamical meaning; feeding it is an invitation to fit the observing history instead of the orbit.\\
These choices are aimed at obtaining a NN model that learns orbits from the relative geometry of the arc rather than from absolute calendar dates. As a bonus, the sub-arc lengths are reused to reconstruct the angular rates inside the training process (Eq.~\eqref{eq:attr_rates}).
By reducing to three the number of observations needed for a prediction, we are able to process objects sooner. This is critical, as most newly discovered NEOs are observed only for a single night and often with just a few measurements.
Another critical architectural choice is the use of two independent linear projection heads for $\rho$ and $\dot\rho$. Therefore, the network outputs two separate scalars $(\rho^*, \dot\rho^*)$ that are trained independently. Formally, we define
\begin{equation}
    \mathbf{y} = (\rho^{*}, \dot\rho^{*})|_{t_{2}}.
\end{equation}

\section{Evaluation metrics}\label{sec:Comparison_metrics}
Here we present all the metrics involved in the comparison of the two NN models investigated in this work, each of which sheds light on a different aspect, or weak point, of these methods.

\subsection{Uncertainty}\label{subsec:uncertainty}
Two different aspects regarding the accuracy of the models' outputs need to be treated. Within the held-out test set, we need a measure of how much their estimations are wrong with respect to the ground truth since we also need to know their confidence in a prediction where no correct answer is available (real deployment scenario).

\subsubsection{Error metrics}
Raw accuracy is reported with four quantities, chosen so that each carries information the others do not. Writing $y$ for the ground-truth, $y^*$ for the prediction and $\sigma_y,\,\sigma_{y^*}$ for their standard deviations over the evaluated set, we consider the mean absolute error $\langle\,|y^* - y|\,\rangle$ (MAE), the median absolute error $\operatorname{med}|y^* - y|$ (medAE), the correlation $\operatorname{corr}(y^*, y)$ and the spread ratio $\sigma_{y^*}/\sigma_y$.\\
These statistical tools, valid for every fitted model measured on a fixed sample, will be used in \Cref{subsec:MLPvsPINN,subsec:classical_comparison} to compare the MLP, the PINN and the classical methods.

\subsubsection{Calibrated prediction intervals}
The dominant error in the VSA regime is the ill-posedness of the inversion itself. The confidence interval that can legitimately be quoted is obtained by \emph{conformal prediction}~\cite{vovk2005algorithmic,lei2018distribution}, which is distribution-free and requires only exchangeability, meaning that the coverage guarantee holds for any underlying model and any data distribution, with no Gaussianity assumption and no requirement that the model be correctly specified.
We introduce the \emph{miscoverage rate} $\beta$, the only free parameter of the construction, chosen as $\beta = 0.32$ and we compute \emph{half-widths} $\hat q$ such that
\begin{equation}
    \mathbb{P}\bigl(\,|\rho^* - \rho| \le \hat q \,\bigr) \;\ge\; 1-\beta \,
    \label{eq:conformal}
\end{equation}
holds inside a \emph{calibration} set. Therefore $\hat q$ represents the smallest quantile below which $1 - \beta = 68\%$ of the absolute residuals $|\rho^* - \rho|$ fall. The uncertainty assigned to a new arc is then $\rho^{*} \pm \hat q$, and the same construction is applied independently to $\dot\rho$. An important design choice adapt the construction to our problem (Mondrian conformal prediction~\cite[\S4.5]{vovk2005algorithmic}): samples are sorted by proper motion $\eta$, cutted at its deciles, and a separate $\hat q$ is computed within each of the ten bins, so that an arc is compared only with arcs of comparable observability. This is necessary since the degeneracy is not uniform across the population, meaning that a single $\hat q$ would be simultaneously too wide for the fast arcs and too narrow for the slow ones. For a newly observed arc the calibration sample is simply the whole held-out test set, meaning one $\hat q$ is computed per bin and assigned to the arc by lookup and it is what the error bars of Figure~\ref{fig:unseen_heldout} report.

\subsection{Energy distribution and Admissible Region}
A complementary diagnostic is whether the predicted state is dynamically admissible at all, i.e.\ whether it corresponds to a bound orbit. Writing the heliocentric position and velocity in terms of $(\rho, \dot\rho)$, the attributable $\mathcal{A}$ and the observer state $(\mathbf{r}_{\oplus}, \mathbf{v}_{\oplus})$, the two-body specific energy admits the closed form~\cite[\S8.1]{milani2010theory}
\begin{equation}
    E_\odot(\rho, \dot\rho) = \tfrac12\bigl(\dot\rho^2 + c_1\dot\rho + c_2\rho^2 + c_3\rho + c_4\bigr) - \frac{\mu_\odot}{\sqrt{\rho^2 + c_5\rho + c_0}} \,,
    \label{eq:helio_energy}
\end{equation}
with the standard coefficients $c_0 = \|\mathbf{r}_{\oplus}\|^2$, $c_1 = 2\,\mathbf{v}_{\oplus}\cdot\hat{\boldsymbol{\rho}}$, $c_2 = \eta^2$, $c_3 = 2(\dot\alpha\,\mathbf{v}_{\oplus}\cdot\partial_\alpha\hat{\boldsymbol{\rho}} + \dot\delta\,\mathbf{v}_{\oplus}\cdot\partial_\delta\hat{\boldsymbol{\rho}})$, $c_4 = \|\mathbf{v}_{\oplus}\|^2$, $c_5 = 2\,\mathbf{r}_{\oplus}\cdot\hat{\boldsymbol{\rho}}$.\\
Boundness alone, however, is only one of the conditions that a physically meaningful state must satisfy. Since the attributable $\mathcal{A}$ is fixed by the observations, the two quantities the network predicts are exactly the two that remain free, and the set of values they may take is the admissible region $\mathcal{D}$ of Milani et al.~\cite{milani2004orbit}, defined by
\begin{equation}
    \mathcal{D} = \left\{(\rho,\dot\rho) \;:\;
    \underbrace{E_\odot(\rho,\dot\rho) \le 0}_{\text{bound to the Sun}},\quad
    \underbrace{E_\oplus(\rho,\dot\rho) \ge 0}_{\text{not bound to the Earth}},\quad
    \underbrace{\rho > R_\oplus}_{\text{physical}}
    \right\},
    \label{eq:admissible_region}
\end{equation}
where the geocentric two-body energy, obtained from the same construction with the Earth as the centre, is
\begin{equation}
    2E_\oplus(\rho,\dot\rho) = \dot\rho^2 + \eta^2\rho^2 - \frac{2\mu_\oplus}{\rho}\,.
    \label{eq:geo_energy}
\end{equation}
We write $P_{\mathrm{AR}}$ for the fraction of arcs whose predicted $(\rho^*,\dot\rho^*)$ falls in $\mathcal{D}$. Its ceiling is not $100\%$: the rates $\dot\alpha,\,\dot\delta$ that complete the orbital state $\mathcal{K}$ come from the three-point fit of Eq.~\eqref{eq:attr_rates} over a single VSA, so the \emph{pseudo-truth} itself can occasionally be inadmissible. On the test set we have $P_{\mathrm{AR}} = 99.4\%$.

\subsection{Southworth--Hawkins distance \texorpdfstring{$D_{\mathrm{SH}}$}{DSH}}
Pointwise $(\rho^{*}, \dot\rho^{*})$ errors and angular residuals do not fully capture the quality of the recovered orbit: in the VSA regime two geometrically distant orbits can fit the same angles with comparably low root mean square errors. We therefore also evaluate the \emph{Southworth--Hawkins} $D$-criterion~\cite{southworth1963statistics} ($D_{\mathrm{SH}}$), an orbit-to-orbit distance computed on the elements $(q, e, i, \Omega, \omega)$ between the NN-seeded orbits and the pseudo-true ones. The criterion is invariant under the choice of reference frame and is defined for elliptic orbits only, so arcs whose predicted or true state is unbound are excluded from the statistics and reported separately.

\section{Results}%
\label{sec:Results}
After the object-disjoint partition of \Cref{subsec:NormScaling}, the held-out test set contains $56{,}773$ arcs belonging to $3{,}865$ distinct objects.
All the numbers reported below come from a single run of the full pipeline on this partition, and every arc is scored by both the MLP and the PINN on exactly the same split. In Table \ref{tab:hyperparams} we summarize the architecture and training hyperparameters used for the PINN model.

\subsection{MLP vs PINN raw comparison}\label{subsec:MLPvsPINN}
\begin{figure}
    \centering
    \includegraphics[width=1\textwidth]{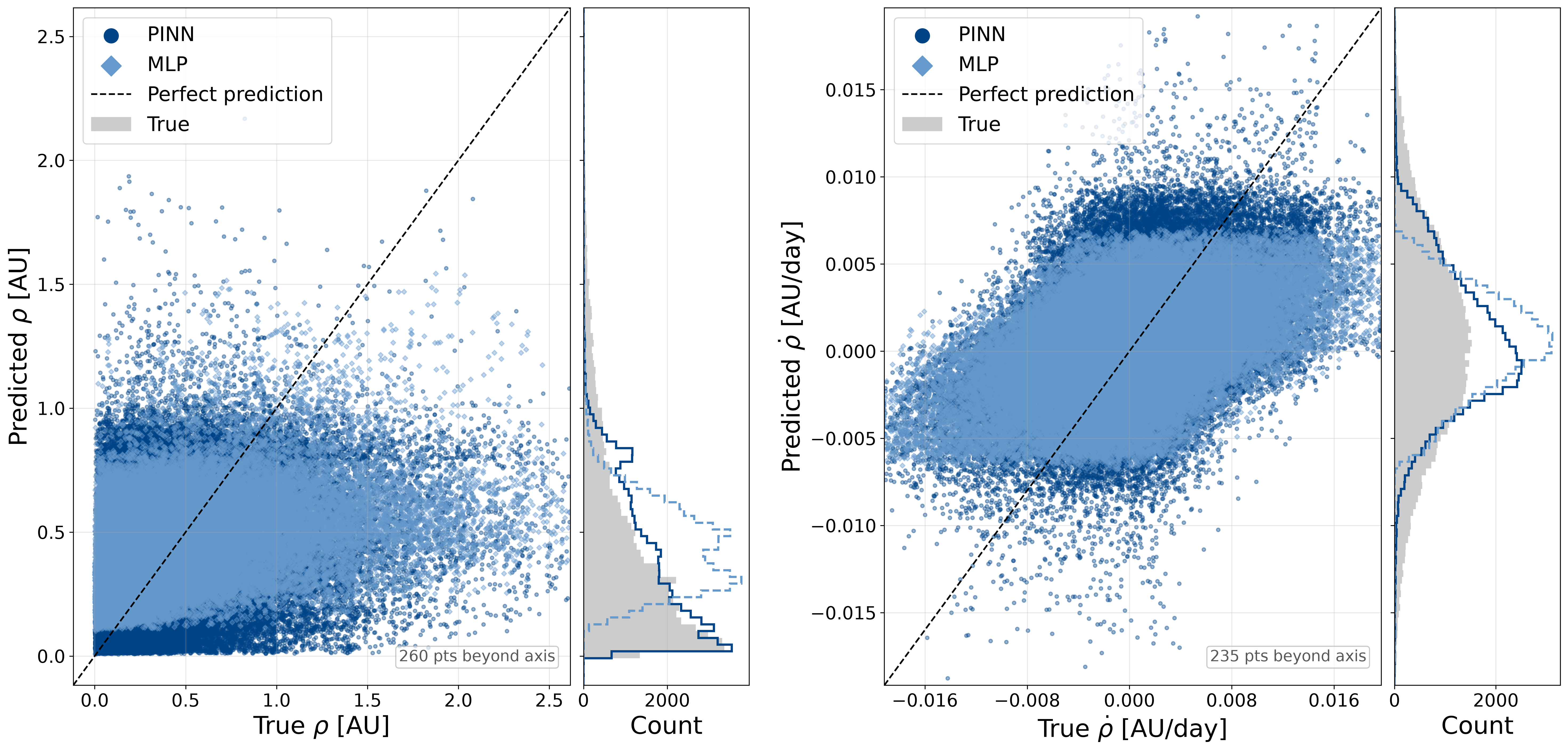}
    \caption{PINN versus MLP baseline on the test set. Neither method reproduces the spread of the ground truth, but the MLP collapses further towards the population median and fails to reproduce the $\rho$ peak of the closest objects.}
    \label{fig:estimation}
\end{figure}

\begin{figure}
    \centering
    \includegraphics[width=1\textwidth]{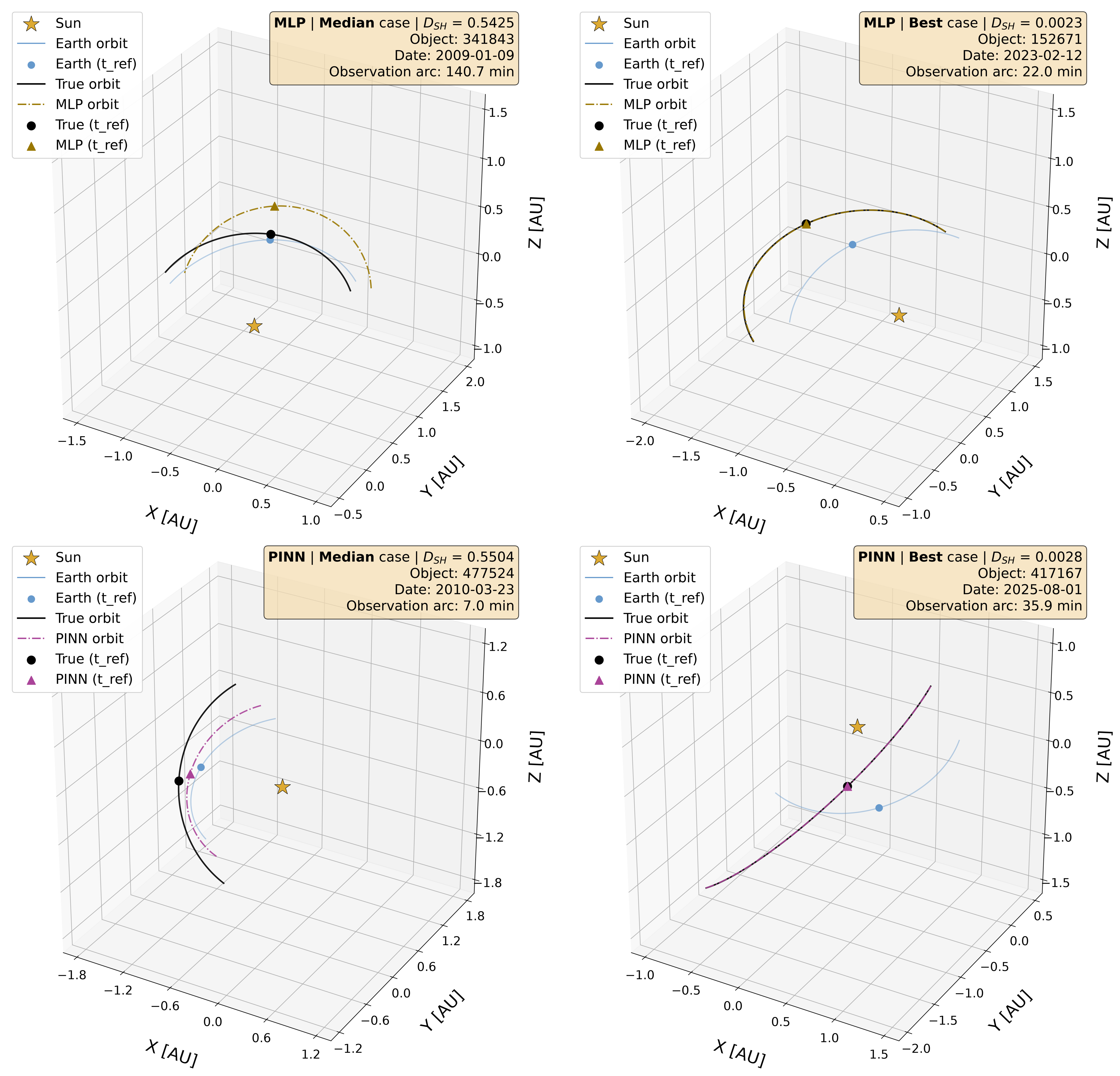}
    \caption{Best and median first-guess orbits for the MLP (top row) and the PINN (bottom row). Each model's test arcs are ranked by the $D_{\mathrm{SH}}$ between the orbit implied by its $(\rho^{*}, \dot\rho^{*})$ and the pseudo-true orbit. The left column shows the median of that ranking, the right column its minimum. Each panel is draws only a part of the Earth, the pseudo-true and the predicted orbit centred on each body's position at the reference epoch $t_{\mathrm{ref}} = t_2$, while Sun is at the origin. The box gives the object, the UTC reference time and the arc's temporal length.}
    \label{fig:dsh_snapshots}
\end{figure}

The MLP baseline and the PINN have been trained through the same infrastructure, on the same splits, with the same input vector of Eq.~\eqref{eq:input_vector} and the same activation. The only deliberate difference between the two models is the objective: the MLP minimises a plain mean squared error on the normalized outputs, while PINN minimises full Eq.~\eqref{eq:total_loss}. It should be kept in mind that, for each sample, the models' predictions rely on information from a single VSA, while the ground truth is computed from the full set of observations available in the NEODyS-2 database.\\
\begin{table}[t]
\centering
\caption{PINN architecture and training hyperparameters.}\label{tab:hyperparams}
\begin{tabular}{@{}llr@{}}
\toprule
\textbf{Category} & \textbf{Parameter} & \textbf{Value} \\
\midrule
\multirow{5}{*}{Architecture}
    & Input dimension & 11 \\
    & Output dimension & 2 (split heads) \\
    & Hidden layers & 5 (widths 512, 256, 128, 64, 32) \\
    & Activation function & Tanh \\
    & Weights initialization & Random Weight Factorization \\
\midrule
\multirow{6}{*}{Training}
    & Optimizer & AdamW (amsgrad) \\
    & Initial learning rate $\gamma_0$ & $2\times10^{-4}$ \\
    & LR schedule & CosineAnnealing ($\gamma_{\min} = 0.1\,\gamma_0$) \\
    & Epochs $N_{\mathrm{ep}}$ & 400 \\
    & Batch size & 512 \\
    & Early stopping patience & 100 epochs \\
\midrule
\multirow{2}{*}{Regularization}
    & Weight decay & $1\times10^{-4}$ \\
    & Gradient clipping & Active, $\|\cdot\|_2 \leq 10.0$ \\
\botrule
\end{tabular}
\end{table}
In Table \ref{tab:MLPvsPINN} we present a comparison of the performance of the PINN and the MLP baseline. We notice that the MLP displays a lower MAE representing not a uniformly better fit but a shorter error tail; on the other hand the PINN, while having a higher MAE, has a lower medAE. This is expected: on an ill-posed inverse problem, a model free to collapse towards the population mean buys accuracy by ceasing to reproduce the real distribution of orbits, at the risk of losing physical interpretability.
\begin{table}
\centering
\caption{Performance comparison between PINN and MLP on the $56{,}773$ arcs in the test set. Best value of each row in bold.}\label{tab:MLPvsPINN}
\begin{tabular}{@{}lrrrr@{}}
\toprule
 & \multicolumn{2}{c}{$\rho$} & \multicolumn{2}{c}{$\dot\rho$} \\
\cmidrule(lr){2-3}\cmidrule(lr){4-5}
 & PINN & MLP & PINN & MLP \\
\midrule
MAE   & $0.304$ [AU]      & $\bm{0.286}$ [AU] & $4.7\times10^{-3}$ [AU/day] & $\bm{4.5\times10^{-3}}$ [AU/day] \\
medAE & $\bm{0.208}$ [AU] & $0.219$ [AU]      & $4.2\times10^{-3}$ [AU/day] & $\bm{4.0\times10^{-3}}$ [AU/day] \\
correlation        & $0.317$           & $\bm{0.391}$      & $0.411$                     & $\bm{0.438}$ \\
spread ratio       & $\bm{0.62}$       & $0.40$            & $\bm{0.63}$                 & $0.46$ \\
\botrule
\end{tabular}
\end{table}
Along with the pointwise errors, we also report the fraction of arcs on which the predicted state is physically admissible and bound to the Sun:
\begin{center}
    PINN: $P_{\mathrm{AR}} = 85.1\%$, \quad MLP: $P_{\mathrm{AR}} = 76.5\%$.
\end{center}
The PINN model is able to produce a physically meaningful output on a larger fraction of arcs than the MLP baseline, which is a direct consequence of the PINN loss function $\mathcal{L}(\theta)$ of Eq.~\eqref{eq:total_loss}.
In Figure~\ref{fig:estimation} the collapse behavior is clearly visible, especially for the MLP baseline and in general for the $\dot\rho^{*}$ distribution. A shrunk estimate is automatically close to a restricted family of orbits and carries no information about the particular object, so point-estimate error alone cannot separate a model that has learned the geometry from one that has learned the prior.
Over the arcs on which a model's first guess and the pseudo-truth are both elliptic ($48{,}105$ for the PINN, $43{,}195$ for the MLP), the median $D_{\mathrm{SH}}$ is $0.550$ for the PINN and $0.543$ for the MLP and, respectively, $7.9\%$ and $8.2\%$ of those arcs have $D_{\mathrm{SH}} < 0.2$ (the threshold generally used to associate two orbits~\cite{southworth1963statistics}). In Figure \ref{fig:dsh_snapshots} we can have a visual representation of this metric. All four predictions showed lie inside the admissible region, which illustrates that admissibility is necessary for a usable first guess but is not a measure of its accuracy. The figure is not a head-to-head comparison: each row comes from its own model's ranking, and on the median arc of either model the other one happens to do considerably better.
The information extrapolated from Table~\ref{tab:MLPvsPINN} is obtained by averaging over a population which is very far from uniform. The range degeneracy of a VSA is severe on slow, distant arcs and comparatively mild on fast, nearby ones, so a single number over all $56{,}773$ arcs mixes regimes that behave differently. This is what motivates the stratified analysis of the following section.

\subsection{MLP vs PINN stratified comparison}\label{subsec:stratified}
\begin{figure}
    \centering
    \includegraphics[width=1\textwidth]{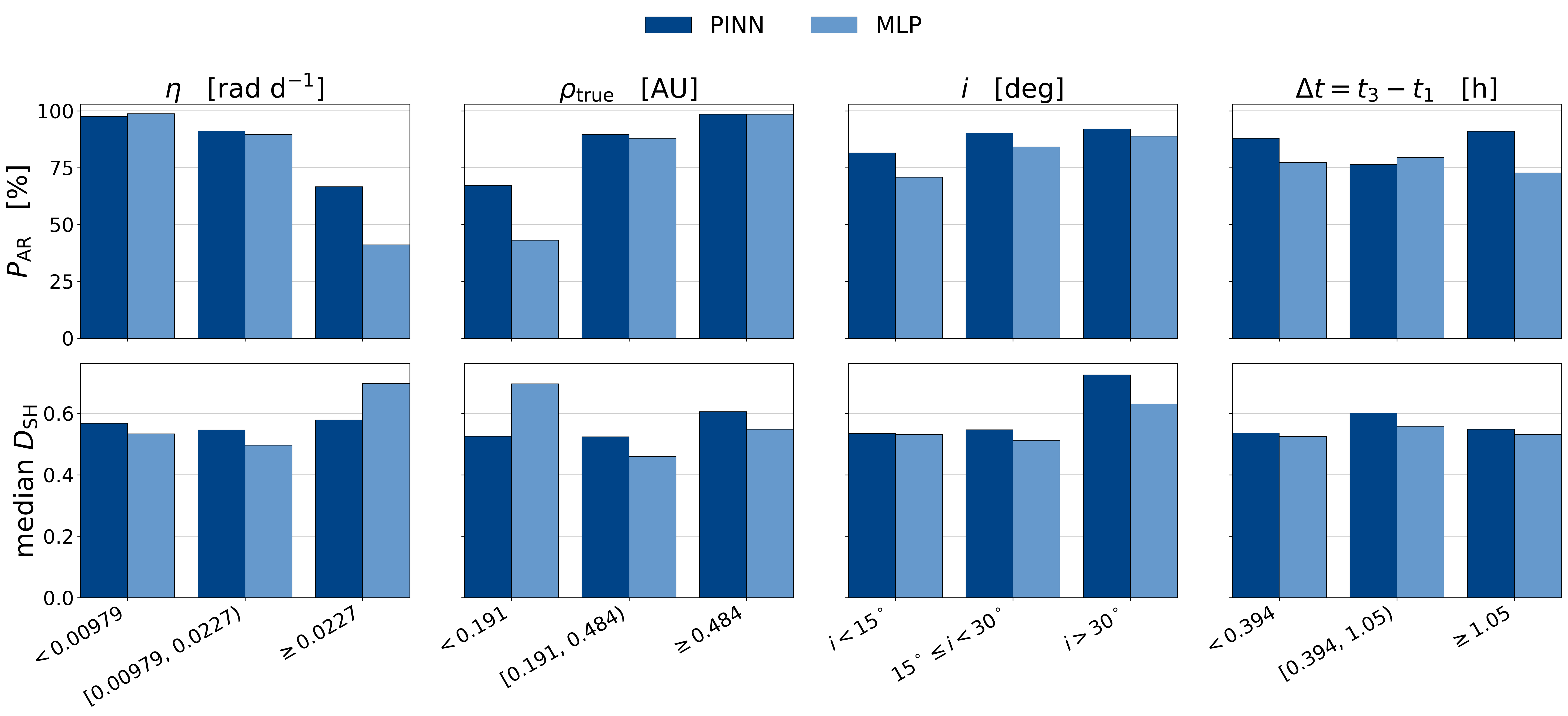}
    \caption{Admissible-region membership $P_{\mathrm{AR}}$ (top row) and median $D_{\mathrm{SH}}$ (bottom row) by object category, over four stratifying axes: proper motion $\eta$, true range $\rho$, inclination $i$ and arc span $\Delta t$. Tick labels give the tercile bounds.}
    \label{fig:stratified_orbits}
\end{figure}

\begin{figure}
    \centering
    \includegraphics[width=1\textwidth]{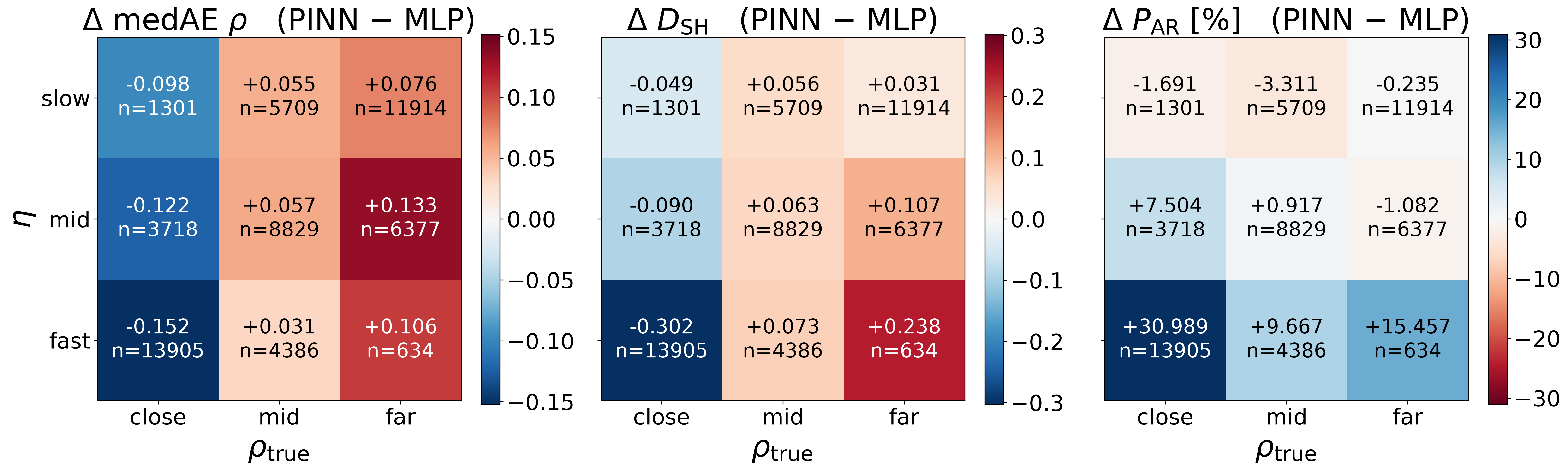}
    \caption{Difference between the two models (PINN $-$ MLP) over the $\eta \times \rho_{\mathrm{true}}$ tercile cross: medAE $\rho$ (left), median $D_{\mathrm{SH}}$ (centre) and $P_{\mathrm{AR}}$ (right). Blue indicates a PINN advantage (a lower medAE and a lower $D_{\mathrm{SH}}$ are better while a higher $P_{\mathrm{AR}}$ is better). Each cell is annotated with its arc count; the $D_{\mathrm{SH}}$ panel is computed on the arcs that are elliptic under both models and the truth.}
    \label{fig:stratified_grid}
\end{figure}

We partition the held-out set into terciles of the proper motion $\eta$ (cut points $slow: \eta < 9.79\times10^{-3}$ and $fast: \eta \geq 2.27\times10^{-2}$~rad/day) and of the true range $\rho$ (cut points $close: \rho_{true} < 0.191$ and $far: \rho_{true} \geq 0.484$~AU), and score both models on $P_{\mathrm{AR}}$ and the median $D_{\mathrm{SH}}$, which asks how far the resulting orbit is from the true one. Because $D_{\mathrm{SH}}$ is undefined for an unbound state, it is always computed on the arcs that are elliptic for both models (and the ground truth).\\
Figure~\ref{fig:stratified_orbits} shows the performance of the NNs on various relevant quantities, each partitioned into three distinct groups representing qualitatively distinct behaviours, similarly to the cases of $\eta$ and $\rho$ presented above. On the slow and far strata the MLP is ahead or level on every metric; on the fast and close strata the ranking inverts sharply. The picture shown in Table~\ref{tab:MLPvsPINN} is therefore not a statement about which model is better, but about which regime dominates the test set by count.
Regarding the $i$ and $\Delta t$ columns, $P_{\mathrm{AR}}$ favours the PINN in every inclination band and on both the shortest and the longest VSAs, the single exception being the middle span tercile. The orbit distance $D_{\mathrm{SH}}$, by contrast, is much flatter across all four axes outside the fast and close strata, which is consistent with the range degeneracy setting a floor that neither model escapes.
Read alone, the first two columns of Figure~\ref{fig:stratified_orbits} are ambiguous, since a fast/slow stratum can be partly a close/far one, and a simple one-dimensional cut cannot expose which axis carries the effect. Figure~\ref{fig:stratified_grid} separates them on the full $\eta\times\rho$ cross.
We note that the PINN's advantage in medAE $\rho$ is present down the entire close column, at every proper motion (from slow to fast), and reverses in every other cell. Its advantage in $P_{\mathrm{AR}}$ is present across the entire fast row, at every range (from close to far).
The two effects meet in the fast/close cell, which carries the largest single-cell differences of the analysis. This is the regime of the recently discovered, nearby objects for which a same-night IOD is operationally worth having, and it is the regime in which the physics-informed objective clearly outperforms the MLP.

\subsection{Comparison with classical IOD methods}\label{subsec:classical_comparison}
\begin{figure}
    \centering
    \includegraphics[width=1\textwidth]{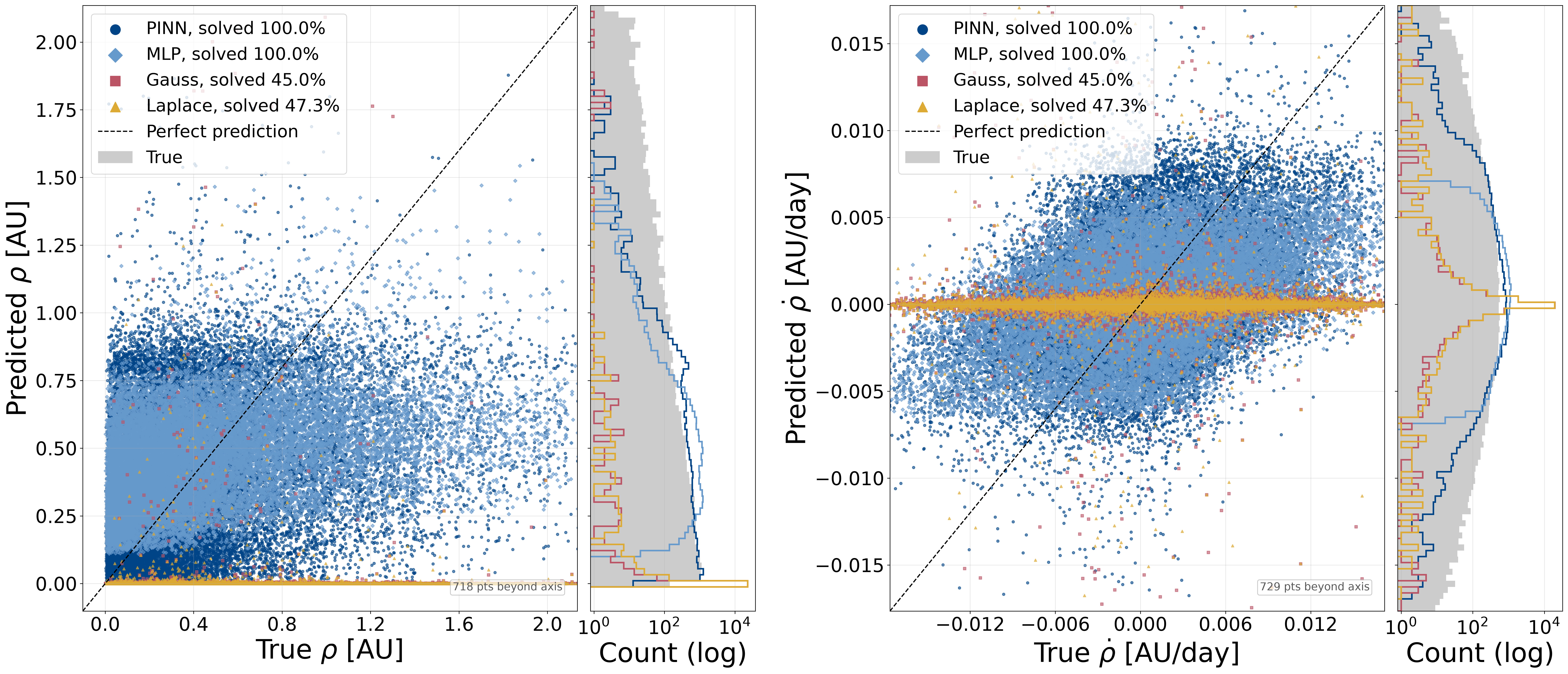}
    \caption{Predicted vs.\ true $(\rho^{*}, \, \dot\rho^{*})$ and estimates' populations histograms for the PINN, the MLP, Gauss and Laplace, restricted to the $23{,}298$ arcs ($41.0\%$ of the held-out set) on which both classical solvers return a solution. The legend reports each method's solve rate over the full held-out set.}
    \label{fig:classic_mutual}
\end{figure}

We solved the classical IOD problem on the same held-out set using both Gauss's and Laplace's method. In Figure~\ref{fig:classic_mutual} we compare the predicted $(\rho^{*}, \,\dot\rho^{*})$ from the four methods over the mutually successful cases.
First of all we notice that the classical methods tends to fail in a large number of cases within the test set: over the $56{,}773$ held-out arcs, Gauss returns a solution on $45.0\%$ and Laplace on $47.3\%$; the two agree on only $23{,}298$ arcs, so Figure~\ref{fig:classic_mutual} is restricted to $41.0\%$ of the test set. The failures are dominated by the root search finding no admissible root in the bracket, which is the numerical signature of the weak curvature discussed in \Cref{subsec:classical}.
Even when successful, the classical estimates carry almost no information about the individual object since they collapse frequently onto the degenerate near-observer root ($92.3\%$ of Gauss's and $88.9\%$ of Laplace's successes return $\rho^{*} < 10^{-3}$~AU, against a true median of $0.310$~AU). This behaviour is also mirrored by their correlation with the truth as we can see in Table~\ref{tab:classical} where we report the comparison on the mutual subset.
\begin{table}
\centering
\caption{The four methods performances on the $23{,}298$ arcs. The rightmost column gives the fraction of the full held-out set on which the method returns a solution at all. True median $\rho = 0.310$~AU.}\label{tab:classical}
\begin{tabular}{@{}lrrrrrr@{}}
\toprule
 & MAE $\rho$ [AU] & medAE $\rho$ [AU] & corr $\rho$ & spread $\rho$ & median $\rho^{*}$ [AU] & solved [\%] \\
\midrule
PINN     & $0.305$ & $0.213$ & $0.315$ & $0.635$ & $0.332$              & $100.0$ \\
MLP      & $0.285$ & $0.219$ & $0.388$ & $0.400$ & $0.442$              & $100.0$ \\
Gauss    & $0.442$ & $0.309$ & $0.015$ & $0.174$ & $5.1\times10^{-5}$   & $45.0$  \\
Laplace  & $0.441$ & $0.309$ & $0.010$ & $0.154$ & $5.6\times10^{-5}$   & $47.3$  \\
\botrule
\end{tabular}
\end{table}
From this analysis we can say that on a single VSA the classical methods either fail outright or return a degenerate root but we cannot claim, for now, that either NN model developed is able to solve the ranging problem satisfactorily despite their superior performances.

\subsection{Performance on new VSAs}\label{subsec:results_uncertainty}
\begin{figure}
    \centering
    \includegraphics[width=0.9\textwidth]{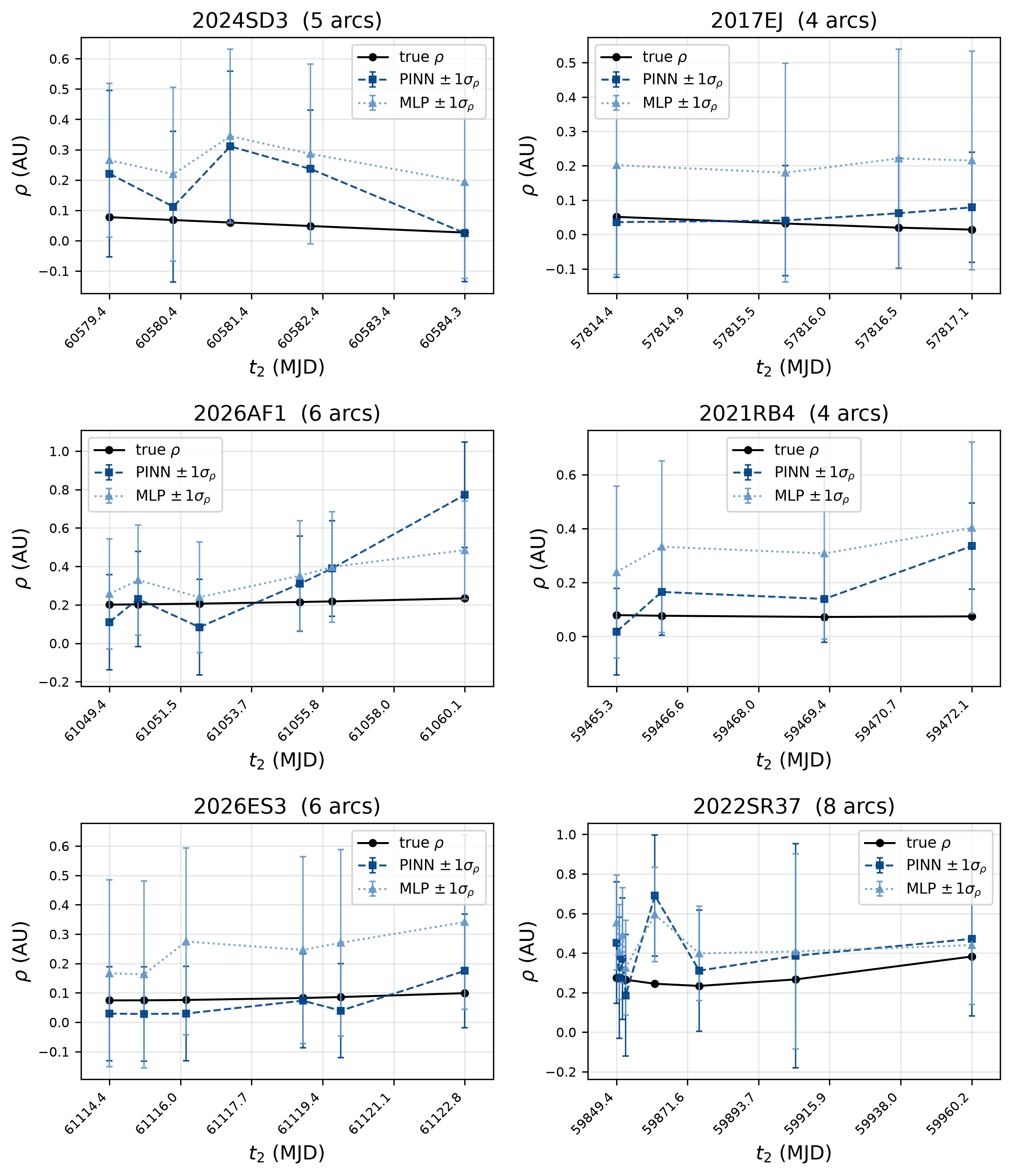}
    \caption{Per-object case study: for six objects drawn at random from a hidden portion of the dataset, the true $\rho$ (solid) and each model's prediction across that object's arcs, ordered by epoch. Error bars are the half-width, with the $\eta$-decile partition of \Cref{subsec:uncertainty} calibrated on the residuals of the whole test set, at the arc's own proper motion.}
    \label{fig:unseen_heldout}
\end{figure}

In Figure~\ref{fig:unseen_heldout} we show six examples of the $\rho^{*}$ estimates produced by our models on VSAs of objects that have been excluded from any split, as a proxy for newly discovered objects. Applying the uncertainty construction of \Cref{subsec:uncertainty} to a single night of observations, processed through the network, localizes a generic NEO along the line of sight to about $\rho \pm 0.31$~AU and $\dot\rho \pm 6.1\times10^{-3}$~AU/day at $1\sigma$.
The cross-conformal band is issued per decile of $\eta$, so it is a distribution and not a single number. The PINN's half-width shrinks monotonically ($0.56$, $0.33$, $0.20$~AU) across the slow, mid and fast terciles, whereas the MLP's is non-monotone ($0.49$, $0.26$, $0.29$~AU). The interval narrows where the geometry is informative and widens where it is not.
The coverage of Eq.~\eqref{eq:conformal} is a guarantee over the population, and the band must not be quoted for an individual category of object.

\section{Conclusions and future work}\label{sec:Conclusion_FutureWork}

In this work, we have presented a novel approach to the ranging problem in IOD from VSAs using Neural Networks and real NEOs observations. The proposed methods exploit data-driven learning and, in the case of the PINN, physical constraints derived from orbital dynamics. These levers compensate for the lack of orbital informations arising from a minimal set of angular observations and enable the estimation of the geocentric $(\rho, \dot\rho)$. Both MLP and PINN estimators are more reliable and accurate than the classical methods in this regime but the intrinsic degenerate nature of the IOD problem is only partially addressed since the mapping from $\mathcal{A}$ to $\mathcal{K}$ is fundamentally non-unique.\\
As expected from an error-optimal estimator on an ill-posed inverse problem, the MLP model better performs in terms of MAE by shrinking towards the population prior. However, once we stratify the test set, the PINN appears to outperform the MLP for the close/fast arcs, which notably are the most critical ones from a planetary defence perspective. The PINN buys feasibility and regime-dependent accuracy at the cost of a longer error tail. The introduction of the Wasserstein loss is primarily motivated for mitigating the mode collapse toward the degenerate range solution. While effective in practice, this component enforces consistency mainly on the distributional level, which may limit its generalization outside the training distribution.
In future work, we are going to investigate alternative approaches to more naturally capture the multi-modal structure of the solution space. Moreover the physical terms in Eq.~\eqref{eq:losses_pde} could be refined to include other planetary perturbation which are comparable in magnitude to the Earth direct gravitational pull, e.g., the Jupiter one.
Finally, the uncertainty reported in Figure~\ref{fig:unseen_heldout} is a marginal prediction interval: the band is valid for the overall dataset population but does not represent a measure of the reliability of the models on newly observed VSAs. We will investigate more advanced methods to quantify this uncertainty. A Differential Corrections convergence analysis using the NNs predictions as initial guesses for the orbit determination is under development.
This will allow us to better assess the quality of the predicted initial orbits and their suitability for further refinement. We are also planning to extend the dataset to include all observed asteroids in the Solar System, not only NEOs.\\
Overall, this work represents a first step toward the development of fast and robust tools for real-time IOD of newly discovered NEOs. Once trained, the models are capable of producing predictions in milliseconds, making it suitable for operational pipelines requiring rapid response.

\backmatter

\section*{Declarations}
\begin{itemize}
\item Funding: This work is supported by the Italian Space Agency (ASI) through the project "Monitoring Asteroids (MonAster)" (grants 2022-33-HH.0 and 2022-33-HH.1).
\item Conflict of interest: not applicable.
\item Ethics approval and consent to participate: not applicable.
\item Consent for publication: not applicable.
\item Data availability: The authors gratefully acknowledge the NEODyS-2 service (operated by University of Pisa, SpaceDys srl and ASI) for providing real and reliable observational data (\url{https://newton.spacedys.com/neodys/}).
\item Materials availability: not applicable.
\item Code availability: available under request.
\item Authors contributions: The first author is responsible for the integrity of the work as a whole, from code development to manuscript completion. Roberto Paoli has contributed to the development of the PINN physical loss terms as well as the classical IOD methods. Riccardo Massidda has contributed to define the machine learning setting of the problem. Giacomo Tommei has contributed with general supervision and guidelines. All authors have contributed to manuscript preparation.
\end{itemize}

\begin{appendices}

\section{\texorpdfstring{$\mathcal{L}_{\mathcal{W}}$}{L\_W} computations}\label{appendix:wasserstein}

The Wasserstein loss is the sum of a global and a conditioned component:
\begin{equation}
    \mathcal{L}_{\mathcal{W}} = \mathcal{L}_{\mathcal{W}}^{\mathrm{global}} + \mathcal{L}_{\mathcal{W}}^{\mathrm{cond}}\,.
    \label{eq:wass_total}
\end{equation}
All quantities are evaluated in the normalized output space of Section~\ref{subsec:NormScaling}, where both channels live in $[-1, 1]$ and enter the quantile and spread terms on an equal footing.

\subsection{Global component}
Let $\sigma_c^{*}, \sigma_c$ and ${\bar y}_c^{*}, \bar y_c$ denote the standard deviation and mean of the predicted and target values for channel $c \in \{\rho,\dot\rho\}$ over the $N$ valid samples, and let $F_c^{-1}(p)$ and ${F_c^{-1}}^{*}(p)$ denote the corresponding $p$-th empirical quantiles. The transport term evaluates Eq.~\eqref{eq:wasserstein_def} on a uniform grid of $m = 51$ levels $q_j = j/(m-1)$. Each quantile is obtained from the sorted sample by differentiable linear interpolation giving
\begin{equation}
    \mathcal{Q}_c = \frac{1}{m}\sum_{j=0}^{m-1}\Bigl|{F_c^{-1}}^{*}(q_j) - F_c^{-1}(q_j)\Bigr|\,,
    \qquad c\in\{\rho,\dot\rho\}\,.
    \label{eq:wass_quantile}
\end{equation}
The grid is a mildly biased estimator of the exact empirical $\mathcal{W}_1$, which is immaterial: the bias is common to every batch and the term is still minimised exactly when the two marginals coincide. Matching the quantile function already constrains the spread, but two auxiliary penalties make the constraint explicit and improve conditioning early in training, when the predicted and target supports barely overlap:
\begin{equation}
    \mathcal{P}_c = \tfrac{1}{2}\bigl(\sigma_c^{*} - \sigma_c\bigr)^2
                  + \tfrac{1}{2}\bigl({\bar y}_c^{*} - \bar y_c\bigr)^2\,,
    \qquad c\in\{\rho,\dot\rho\}\,.
    \label{eq:wass_spread}
\end{equation}
Tail penalties preserve the support of both channels via one-sided quadratic hinges on the $10$th and $90$th percentiles:
\begin{equation}
    \mathcal{T}_c = \Bigl[\mathrm{ReLU}\bigl(F_c^{-1}(0.1) - {F_c^{*}}^{-1}(0.1)\bigr)\Bigr]^2
    + \Bigl[\mathrm{ReLU}\bigl(F_c^{-1}(0.9) - {F_c^{*}}^{-1}(0.9)\bigr)\Bigr]^2, \quad c\in\{\rho,\dot\rho\}.
    \label{eq:wass_tail}
\end{equation}
Each hinge is one-sided, meaning it penalizes the predicted quantile falling inside its target counterpart, counteracting the inward mass drift characteristic of the mode-collapse failure, while leaving over-coverage unpenalized. The global component is therefore
\begin{equation}
    \mathcal{L}_{\mathcal{W}}^{\mathrm{global}} = \sum_{c\,\in\,\{\rho,\dot\rho\}}\bigl(\mathcal{Q}_c + \mathcal{P}_c + \mathcal{T}_c\bigr)\,.
    \label{eq:wass_global}
\end{equation}

\subsection{Conditioned component}
The proper motion at the reference epoch $t_2$ is
\begin{equation}
    \eta_i = \sqrt{\dot\alpha_i^2\cos^2\!\delta_i + \dot\delta_i^2}\,,
    \label{eq:proper_motion}
\end{equation}
where $(\dot\alpha_i, \dot\delta_i)$ are the attributable angular rates (Eq.~\eqref{eq:attr_rates}) and $\delta_i$ is the declination at $t_2$. In practice the split is performed on $\eta_i^2$: samples are labelled \emph{fast} if $\eta_i^2 \geq \tilde{\eta}^2$ (where $\tilde{\eta}$ is the batch median) or \emph{slow} otherwise, and \emph{north}/\emph{south} by $\mathrm{sgn}(\delta_i)$, giving four bins $\mathcal{B}_b$. Within each bin holding at least $8$ samples the same $\mathcal{W}_1$ matching of Eq.~\eqref{eq:wass_quantile} is applied to $\rho$ alone, producing a per-bin loss
\begin{equation}
    \ell_b = \frac{1}{|\mathcal{B}_b|}\sum_{k=1}^{|\mathcal{B}_b|}\bigl|\rho^{*(b)}_{(k)} - \rho^{(b)}_{(k)}\bigr|\,,
    \qquad
    \mathcal{L}_{\mathcal{W}}^{\mathrm{bins}} = \frac{\sum_b |\mathcal{B}_b|\,\ell_b}{\sum_b |\mathcal{B}_b|}\,,
    \label{eq:wass_bin}
\end{equation}
where $\rho^{*(b)}_{(k)}$ and $\rho^{(b)}_{(k)}$ are the $k$-th order statistics within the bin. Only $\rho$ is treated this way: the range rate has no comparable systematic dependence on proper motion, and is left to the global component. An ordering constraint then enforces the physical prior that fast-moving objects are closer,
\begin{equation}
    \mathcal{C}_{\mathrm{ord}} = \Bigl[\mathrm{ReLU}\bigl(\bar\rho^*_{\mathrm{fast}} - \bar\rho^*_{\mathrm{slow}}\bigr)\Bigr]^2,
    \qquad
    \mathcal{L}_{\mathcal{W}}^{\mathrm{cond}} = \mathcal{L}_{\mathcal{W}}^{\mathrm{bins}} + \mathcal{C}_{\mathrm{ord}}\,,
    \label{eq:wass_order}
\end{equation}
where $\bar\rho^*_{\mathrm{fast}}$ and $\bar\rho^*_{\mathrm{slow}}$ are the mean predicted ranges over the respective half-populations, evaluated only when both have $\geq 8$ samples.

\end{appendices}

\bibliography{sn-bibliography}%

\end{document}